\documentclass[prl,longbibliography,twocolumn,superscriptaddress,english,showpacs]{revtex4-1} 

\usepackage[english]{babel}
\usepackage[dvipsnames]{xcolor} 
\usepackage{graphicx} 
\usepackage{amsmath} 
\usepackage{amssymb}
\usepackage{verbatim}
\usepackage{ctable}
\definecolor{darkblue}{rgb}{0.1,0.2,0.6} \definecolor{darkred}{rgb}{0.8,0.1,0.2}
\usepackage[colorlinks,citecolor=darkblue,linkcolor=darkred,urlcolor=darkblue]{hyperref}
\usepackage[all]{hypcap} 

\newcommand{\Z}{{\mathbb Z}}
\newcommand{\R}{{\mathbb R}}
\newcommand{\dd}{{\mathrm d}}

\newcommand{\ed}{{\mathrm e}}

\newtheorem{claim}{Claim}

\begin{document}

\title{Pre-thermalization in a classical phonon field: slow relaxation of the number of phonons} 
\author{Fran\c{c}ois Huveneers}
\affiliation{Universit\'e Paris Dauphine-PSL, CEREMADE, 
Place du Mar\'echal de Tassigny, 75016 Paris, France}
\email{huveneers@ceremade.dauphine.fr}

\author{Jani Lukkarinen}
\affiliation{University of Helsinki,Department of Mathematics and Statistics,
P.O. Box 68, FI-00014 Helsingin yliopisto, Finland}
\email{jani.lukkarinen@helsinki.fi}

\date{\today}

\begin{abstract} 
\noindent
We investigate the emergence of an astonishingly long pre-thermal plateau in a classical phonon field, here a harmonic chain with on-site pinning.
Integrability is broken by a weak anharmonic on-site potential with strength $\lambda$. 
In the small $\lambda$ limit, the approach to equilibrium of a translation invariant initial state is described by kinetic theory. 
However, when the phonon band becomes narrow, we find that the (non-conserved) number of phonons relaxes on much longer time scales than kinetic.
We establish rigorous bounds on the relaxation time, and develop a theory that yields exact predictions for the dissipation rate in the limit $\lambda \to 0$. 
We compare the theoretical predictions with data from molecular dynamics simulations and find good agreement. 
Our work shows how classical systems may exhibit phenomena which at the first glance appear to require quantization.
\end{abstract}

\maketitle

\noindent
\textbf{Introduction ---}
Thermalization is one of the most commonly encountered physical phenomena and yet, it still remains poorly understood. 
Several materials have been found where the approach to equilibrium can be drastically slowed down or even suppressed:
Anderson insulators \cite{anderson_1958},
many-body localized chains \cite{gornyi_et_al_2005,basko_et_al_2006}, 
ergodic systems featuring many-body scars \cite{turner_et_al_18}, 
quantum glasses \cite{kagan_maksimov_1984}, 
Fermi-Pasta-Ulam-Tsingou chains \cite{fermi_et_al_1955},
classical non-linear disordered lattices \cite{basko_2011}, etc.
Moreover, some systems with otherwise good ergodic properties, may feature extensive pseudo-conserved quantities that relax only on very long time scales
\cite{dalessio_polkovnikov_2013,dalessio_rigol_2014,lazarides_et_al_2014,abanin_et_al_2015,abanin_et_al_prb_2017,abanin_et_al_cmp_2017,mori_et_al_2016,
sensarma_et_al_2010,vajna_et_al_2018,de_roeck_verreet_2019,
carati_maiocchi_2012,giorgilli_et_al_2015,howell_et_al_2019}.  
The period during which this quantity stays approximately conserved provides an example of pre-thermal state, 
that may host a lot of fascinating physical phenomena
\cite{else_et_al_2017,else_et_al_2019,else_fendley_et_al_2017,lindner_et_al_2017,martin_et_al_2017}.

\begin{figure}[h]
    \centering
   	\includegraphics[draft=false,height = 3cm,width = 7.5cm]{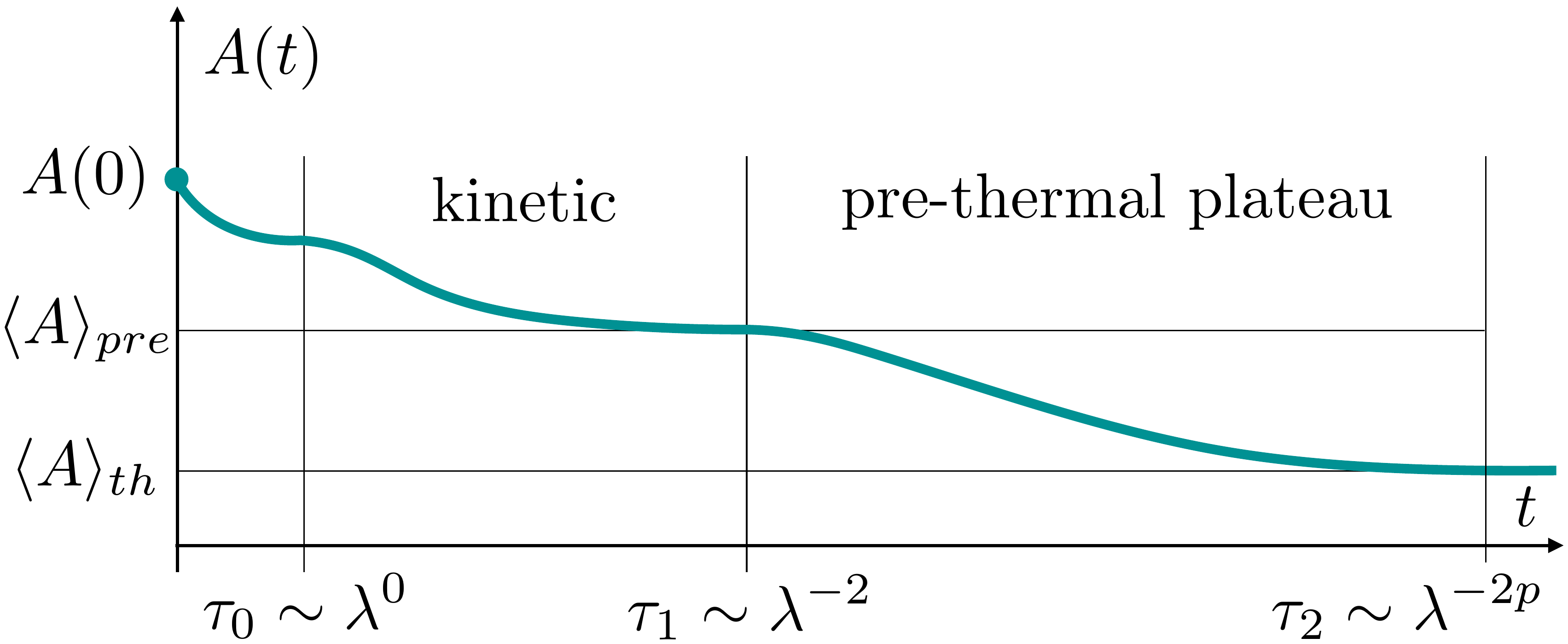}
    \caption{Expected time evolution of a local observable $A(t)$. }
    \label{fig: overall picture}
\end{figure}

In this letter, we investigate a classical many-body Hamiltonian $H = H_0 + \lambda V$ in the limit $\lambda \to 0$, 
where $H_0$ is integrable (a harmonic lattice or free phonon field) and $V$ breaks integrability (an on-site anharmonic potential).
If the system is started in a translation invariant state, 
its state evolves swiftly to the generalized Gibbs ensemble (GGE), characterized by all the conserved quantities of $H_0$
\cite{dudnikova_et_al_2003,rigol_et_al_2007,vidmar_rigol_2016,essler_fagotti_2016}.
Next, according to kinetic theory and the Boltzmann-Peierls equation, it approaches equilibrium in a time $\tau_1$ with $\tau_1\sim\lambda^{-2}$ for $\lambda \to 0$ 
\cite{peierls_1929,spohn_2006,mendl_et_al_2016,lukkarinen_2016}.
However, if the phonon band is sufficiently narrow, the (non-conserved) number of phonons is preserved by kinetic processes \cite{spohn_2006b,mendl_et_al_2016}, 
and only a pre-thermal plateau is reached on kinetic time scales.
As our analysis reveals, the proper equilibrium is only reached after a longer time $\tau_2$, scaling as $\tau_2\sim\lambda^{-2p}$ for some $p\ge 1$. 
See Fig.~\ref{fig: overall picture} for a summary of the above process.

The presence of an almost conserved quantity (or adiabatic invariant) for the Hamiltonian $H$ studied in this letter, 
has been realized in \cite{carati_maiocchi_2012,giorgilli_et_al_2015}.
Here, we first provide rigorous quantitative bounds on the dissipation of the number of phonons, 
see Claims~\ref{cl: new variables} and \ref{cl: dressed quantity} below. 
In addition, we connect them to the slow dissipation of some quantized fields \cite{sensarma_et_al_2010,abanin_et_al_cmp_2017},
a phenomenon that seemed to require quantization. 
Second, we provide a theory to compute the dissipation rate of the pseudo-conserved quantity, which we are able to back with numerical results. 
General predictions for the rate have been derived in \cite{mallayya_et_al_2019} for quantum systems, 
see also \cite{lenarcic_et_al_2018,lange_et_al_2018,reimann_dabelow_2019}.
However, we notice that our system is classical 
and that an extra time-scale is present since relaxation to the pre-thermal plateau involves kinetic processes.

\noindent
\textbf{Model --}
Let the Hamiltonian $H$ be given by
\begin{equation}\label{eq: hamiltonian}
	H 
	= \sum_{x \in \Z}
	\frac{p_x^2}{2} + \frac{\omega_0^2}{2}q_x^2 - \frac{\omega_0^2 \delta}{2} (q_{x-1}q_x + q_x q_{x+1})  + \frac{\lambda}{r} q_x^r
\end{equation}
with $r>2$ an even integer (below we focus on $r=4,6$), and $\omega_0>0$ a characteristic frequency of the system.
The dynamics is classical: $\dot q_x = p_x$ and 
\begin{equation*}
	\dot p_x =
	-(1 - 2 \delta) \omega_0^2 q_x + \delta \omega_0^2 (q_{x+1} - 2 q_x + q_x) - \lambda q_x^{r-1} 
\end{equation*}
where $\dot x$ denotes the time derivative of $x$.
Stability imposes $\lambda \ge 0$ and $0 \le \delta \le 0.5$. 
The chain is uncoupled for $\delta = 0$ and unpinned for $\delta = 0.5$, and we restrict our attention to $0 < \delta < 0.5$. 
If needed, we may obviously restrict the summation in eq.~\eqref{eq: hamiltonian} to $1 \le x \le L$ for a length $L$, 
and consider the limit $L \to \infty$ only afterwards. 

For $\lambda = 0$, the chain is harmonic. 
For a pseudo-momentum $k$ in the Brillouin zone $\mathrm{BZ}= {]}{-0.5},0.5]$, let the phonon mode $a(k):=a^{-}(k)$ be defined by 
\begin{equation}\label{eq: phonon field}
	a^\pm (k) = \frac{1}{\sqrt{2}}\left( \omega^{1/2}(k) \hat q(k) \mp i \frac{1}{\omega^{1/2}(k)} \hat p(k) \right) 
\end{equation}
with the dispersion relation 
\begin{equation}\label{eq: dispersion relation}
	\omega (k) = \omega_0 \left(1 - 2 \delta \cos (2 \pi k) \right)^{1/2}
\end{equation}
and with $\hat f$ the Fourier transform of $f$, defined by 
$\hat f (k) = \sum_{x\in \Z} f(x) \ed^{- 2 i \pi kx}$.
We identify $n(k) = |a(k)|^2$ with the number of phonons with pseudo-momentum $k$ and we denote 
the total number of phonons by $N_0 = \int_{\mathrm{BZ}} \dd k\, n(k)$.
From the analyticity of $\omega(k)$ in eq.~\eqref{eq: dispersion relation}, we deduce that $N_0$ is quasi-local, i.e.\@ 
$$
	N_0 = \sum_{x,y} K_{qq}(x-y) q_xq_y + K_{qp}(x-y) q_x p_y + K_{pp}(x-y) p_xp_y
$$
where the kernels $K_{\star} (z)$ decay exponentially with $|z|$.

When $\lambda > 0$, the Hamiltonian \eqref{eq: hamiltonian} can be written as $H = H_0 + \lambda V$ with 
$H_0 = \int_{\mathrm{BZ}} \dd k \, \omega(k) n(k)$ and
\begin{equation}\label{eq: V}
	V = \frac{1}{r 2^{r/2}} \int_{(\mathrm{BZ})^r} \hspace{-0.6cm}\dd k_1 \dots \dd k_r  \frac{\delta (k_1 + \dots + k_r)}{(\omega_1 \dots \omega_r)^{1/2}} 
	\sum_{\sigma_j = \pm} a^{\sigma_1}_1 \dots a^{\sigma_r}_r \, 
\end{equation}
with the notations $\omega_j = \omega(k_j)$ and $a_j^{\sigma_j} = a^{\sigma_j}(k_j)$.
From eq.~\eqref{eq: V} and the Poisson bracket rule $\lbrace a^\sigma_k, a^{\sigma'}_{k'} \rbrace = i \sigma \delta_{\sigma+\sigma'} \delta (k + k')$, 
we deduce that $N_0$ is not conserved, i.e. $\lbrace H, N_0 \rbrace \ne 0$, 
where $\lbrace \cdot, \cdot \rbrace$ denotes the Poisson bracket, due to the terms with $\sum_{j=1}^r \sigma_j \ne 0$.

\bigskip
\noindent
\textbf{Pseudo-conservation of $N_0$ ---}
To compute the time-scales on which $N_0$ gets dissipated, let us first assume that the system is quantized, 
i.e.\@ that $a^{\pm}$ are creation/annihilation operators for bososns.
Later on, we will see how the conclusions carry over to the classical system.   
In this case, $N_0$ has integer spectrum. 
In the limit $\lambda \to 0$, only resonant processes, preserving the bare energy $H_0$, do effectively destroy the conservation of $N_0$. 
Therefore, in first order in $\lambda$,
the process of creating two phonons (since $r$ is even, it is not possible to create only one phonon)  must satisfy the constraints
\begin{eqnarray}
	&&\omega_1 + \dots + \omega_{n/2 + 1} - \omega_{n/2 + 2} - \dots - \omega_n \; = \; 0,\nonumber\\
	&&k_1 + \dots + k_n = 0\label{eq: resonant constraints}
\end{eqnarray}
where $n=r$ (we will later consider $n\ne r$ when dealing with higher order processes),
and where the second constraint is taken modulo 1 and stems from translation invariance, cfr.\@ eq.~\eqref{eq: V}.
Since the width of the dispersion relation in \eqref{eq: dispersion relation} scales as $2 \omega_0 \delta$ for small $\delta$, 
larger and larger values of $r$ are needed to satisfy the constraints in eq.~\eqref{eq: resonant constraints}. 
Thus, for given $n\ge 2$ even, there exists $\delta_c (n)$ such that eq.~\eqref{eq: resonant constraints} only has solutions for $\delta \ge \delta_c (n)$: 
$\delta_c (2) = \delta_c(4) = 0.5$ (exact), $\delta_c (6) = 0.3$ (exact), $\delta_c (8) \simeq 0.25$ (numerical), 
and in general $\delta_c (n) \sim 2/n$ for $n\to \infty$, see the Supplemental Material (SM). 
From now on, we will assume that $\delta$ is such that $\delta \ne \delta_c (n)$ for any $n$, leaving these exceptional cases for further studies.

Higher order processes need obviously to be taken into account in the case $\delta < \delta_c(r)$. 
The analysis detailed in the SM yields that a process of order $\lambda^p$
involves the creation/annihilation of $n = p (r-2) + 2$ phonons\footnote{This 
value may also be recovered from a simple power counting argument: 
Let $\mathrm{ad}_H =\{H, \cdot \}$ ($= -i[ H, \cdot]$ for a quantum system with $\hbar=1$) and expand $\ed^{-t\mathrm{ad}_H} N_0$ at order $p$; 
this yields $(-t)^p\mathrm{ad}_H^p (N_0) / p!$ which is a polynomial of order $n$ in $a^\pm$, with $n$ as given.},
and the above analysis caries over with this new value for $n$.
Given $r$ and $\delta$, we may now determine the smallest $p$ so that a process of order $\lambda^p$ destroys $N_0$ effectively: 
$p\ge 1$ is the only integer such that 
\begin{equation}\label{eq: p value}
	\delta_c (p(r-2)+2) < \delta < \delta_c ((p-1)(r-2)+2).
\end{equation}
Explicit results are gathered on Table~\ref{table: value of p}: 

\begin{table}[h]
	\centering
	\begin{tabular}{cccc}
		\specialrule{.1em}{0em}{0em}
		&\vline& $r=4$ \vline & $r=6$ \\
		\hline
		$0.3 < \delta < 0.5$ &\vline & $p=2$ \vline & $p=1$ \\
		$0.25 < \delta < 0.3$ &\vline & $p=3$ \vline & $p=2$ \\
		\hline
		\multicolumn{4}{c}{$\dots$}\\
		\hline
		$\delta \to 0$ &\vline& $p \sim 1/\delta$ \vline & $p \sim 1 / 2 \delta$\\
		\specialrule{0.1em}{0em}{0em}
	\end{tabular}
	\centering
   	\caption{Order of the processes $(\lambda^p)$ destroying effectively the conservation of $N_0$, for $r=4,6$ in eq.~\eqref{eq: V}. } 
    \label{table: value of p}
\end{table}  

We observe that, even though quantitative statements are obviously model dependent, 
and in particular the threshold values $\delta_c$ depend on the specific form of the dispersion relation $\omega(k)$ in eq.~\eqref{eq: dispersion relation}, 
the conclusion that $p \sim c/\delta$ as $\delta \to 0$ is generic (for a polynomial interaction $V$), 
following from the fact that the width of the band in $\omega(k)$ decays as $\delta$ for $\delta \to 0$. 

The above analysis may be turned into rigorous results, using the formalism developed in \cite{abanin_et_al_cmp_2017}
that can be straightforwardly adapted to a classical system through the canonical replacement $- i [H , \cdot]$ by $\lbrace H, \cdot\rbrace$, see SM and below.
The key observation to proceed is that, even though $N_0$ is no longer quantized, the spectrum of $\mathrm{ad}_{N_0} = \lbrace N_0,\cdot\rbrace$ is: 
\begin{equation*}
	\lbrace N_0 , a_1^{\sigma_1} \dots a_m^{\sigma_m} \rbrace = i(\sigma_1 + \dots + \sigma_m) a_1^{\sigma_1} \dots a_m^{\sigma_m}.
\end{equation*}
Hence, it acts formally in the same way as the quantum super-operator $-i [N_0,\cdot]$, and this is eventually what matters. 

Let us fix $r$ and $\delta$ such that $p>1$, and let us first express a result in a formulation directly inspired from \cite{abanin_et_al_cmp_2017}: 
There exists a canonical change of variables, bringing $H$ into $\tilde H$, so that $\lbrace \tilde H , N_0 \rbrace = \mathcal O (\lambda^p)$, 
where the term $\mathcal O (\lambda^p)$ is extensive. 
To formulate a precise claim, we will consider extensive observables $\varphi_m$ of the form
\begin{multline}\label{eq: order n observable}
	\varphi_m
	= 
	\int_{(\mathrm{BZ})^m} \dd k_1 \dots \dd k_n \delta(k_1 + \dots + k_m) \\
	\sum_{\sigma_j = \pm} \hat\varphi_m(k_1,\dots,k_n,\sigma_1, \dots ,\sigma_m)a_{1}^{\sigma_1} \dots a_m^{\sigma_m}
\end{multline}
where $m\ge 2$ and where $\hat\varphi_m$ is analytic in $(k_1,\dots ,k_m)$, ensuring that $\varphi_m$ is a sum of quasi-local observables. 
A function $\varphi$ will simply be said to be a polynomial (of order $m$)
if it is of the form $\varphi = \sum_{k=2}^m \varphi_k$ with $\varphi_k$ as in eq.~\eqref{eq: order n observable}. 
The following claim is shown in the SM:
\begin{claim}\label{cl: new variables}
Let $L$ be finite and let us assume periodic boundary conditions. 
For $|\lambda|$ small enough, 
there exists a polynomial $G = \lambda\sum_{n=1}^{p-1}\lambda^{n-1}G_n$, 
such that $\tilde H := \ed^{- \lbrace G,\cdot\rbrace}H$ is a well defined real analytic function in a neighborhood of the origin in $\R^{2L}$, 
and such that $\lbrace \tilde H, N_0 \rbrace = \lambda^p\sum_{n=p}^\infty \lambda^{n-p}\mathcal J_n$, 
where $\mathcal{J}_n$ are polynomials and where the expansion converges to an analytic function in a neighborhood of the origin in $\R^{2L}$.
\end{claim}
Claim~\ref{cl: new variables} provides a good way to think about the phenomenon but does not yield as such a very powerful result in the thermodynamic limit: 
The radius of convergence of $\lambda$ may shrink as $L \to \infty$ 
(even though we are interested in the regime $\lambda \to 0$, the limit $L \to \infty$ needs to be taken before the limit $\lambda \to 0$). 
As a way out, we may undo the above transformation to obtain a dressed number of phonons, $\ed^{\lbrace G , \cdot \rbrace } N_0$, and then truncate its expansion at order $p-1$. 
Doing so, we end up with a well defined pseudo-conserved quantity $N$ in the thermodynamic limit, see the SM for details:
\begin{claim}\label{cl: dressed quantity}
Let now the chain be defined on the full lattice $\Z$. 
There exists a quantity $N = N_0 + \lambda \sum_{n=1}^{p-1} \lambda^{n-1} N_n$, 
where $N_n$ are polynomials of order $nr - 2(n-1)$ for $1 \le n \le p-1$, 
such that $\lbrace H, N \rbrace = \lambda^p \lbrace V,N_{p-1}\rbrace$. 
\end{claim}

These claims furnish upper bounds on the dissipation rate of $N$, but the determination of this rate requires the knowledge of the instantaneous state of the system. 
To proceed further, we will invoke additional assumptions, and leave mathematical rigor behind.

\begin{figure}[h]
    \centering
   	\includegraphics[draft=false,height = 5cm,width = 7.5cm]{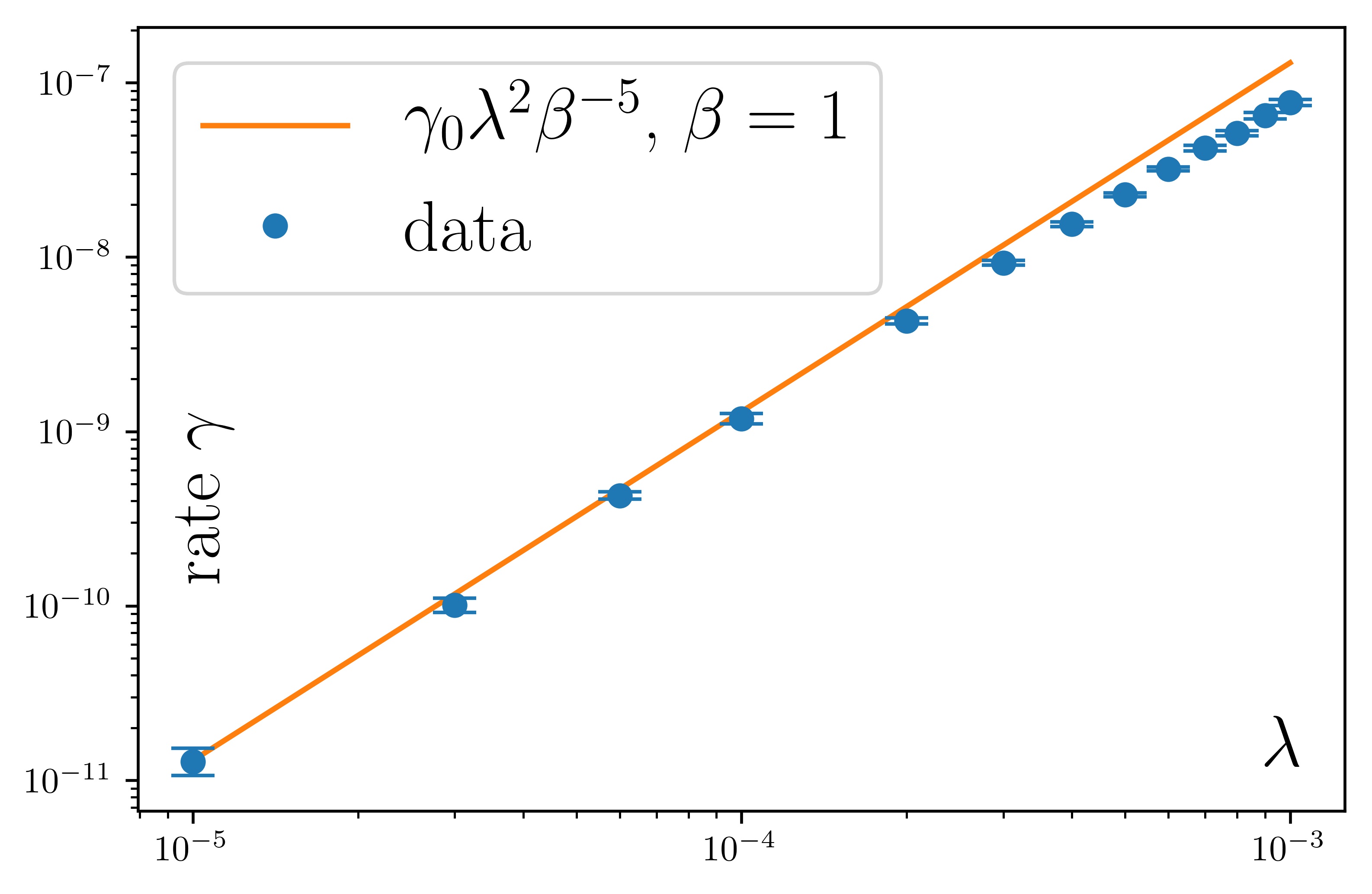}
	\includegraphics[draft=false,height = 5cm,width = 7.5cm]{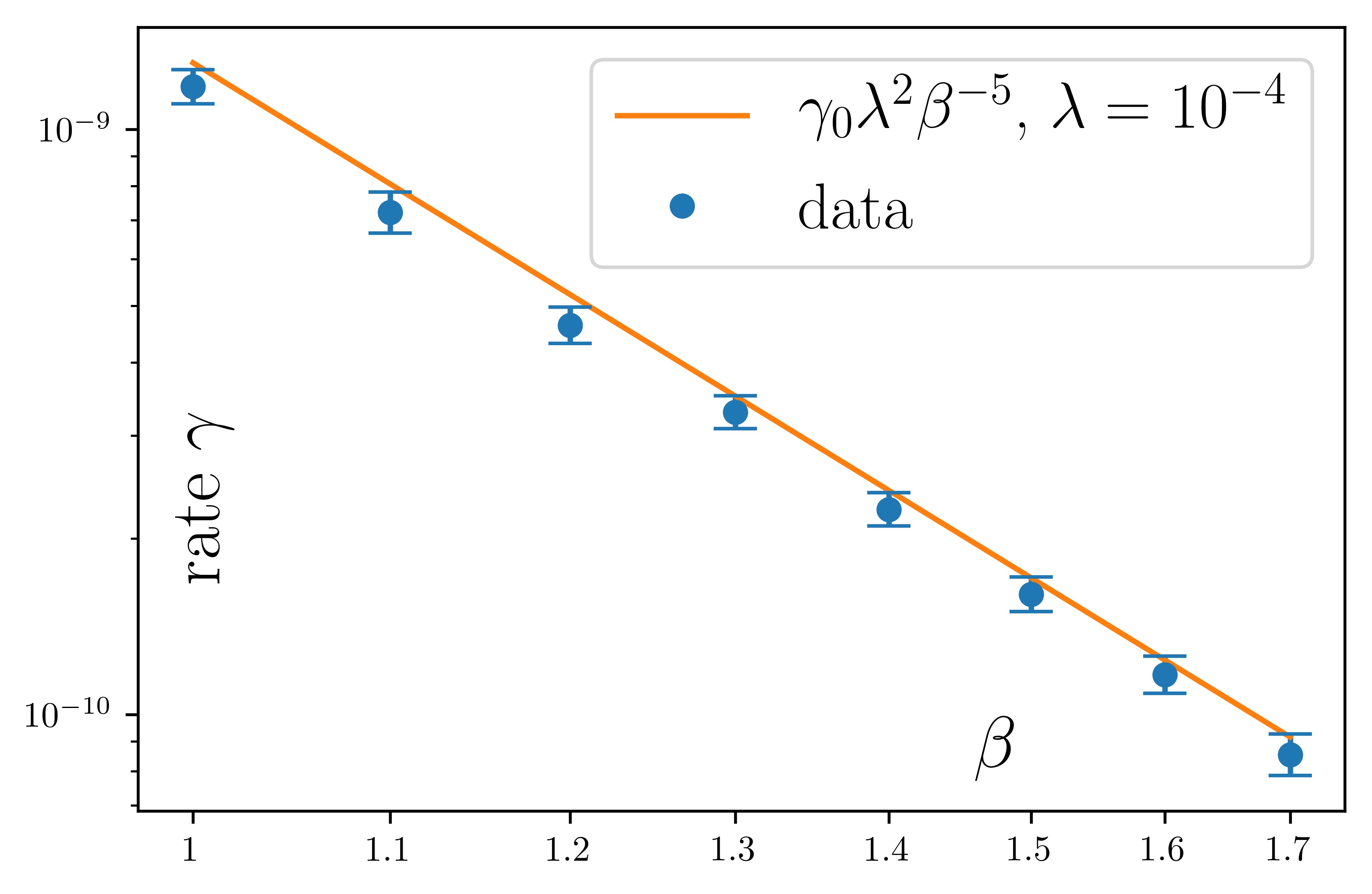}
    \caption{$\gamma$ as a function of $\lambda$ (upper panel) and $\beta$ (lower panel) for $r=6$ and $\delta = 0.35$. 
    Kinetic theory, i.e.\@ eq.~\eqref{eq: rate p=1}, predicts $\gamma  = \gamma_0 \lambda^2 \beta^{-5}$ with $\gamma_0\simeq 0.13$ for $\lambda \to 0$.} 
    \label{fig: r=6 delta=0.35}
\end{figure}

\bigskip
\noindent
\textbf{Evaluation of the decay rate ---}
For $p=1$, $N = N_0$. 
The dissipation rate $\gamma$ of the density $N_0/L$ in an instantaneous state $\rho$ and in the infinite volume limit is given by $\gamma = \langle J \rangle_\rho/\delta (0)$, 
with a flux $J = \lambda \mathcal J = \lambda \lbrace V,N_0 \rbrace$ and where "$\delta (0)$"
stands for the infinite volume, corresponding to $L$ in a chain of finite length.
If the system is prepared in a translation invariant state with zero average, after a short transient time, 
it evolves towards a GGE characterized by a Wigner function $W$, 
i.e.\@ a Gaussian state $\ed^{- \int_{\mathrm{BZ}}\dd k n(k)/W(k)}/Z$. 
Usual kinetic theory yields the following expression for the rate $\gamma (W)$:
Given a function $\varphi = \int_{\mathrm{BZ}} \dd k \hat\varphi (k) n(k)$, let $J_\varphi = \lambda \mathcal J_\varphi = \lbrace H,\varphi\rbrace = \lambda \lbrace V,\varphi\rbrace$, then
\begin{equation}\label{eq: rate p=1}
	\gamma(W)  = \frac{\lambda^2}{\delta (0)} 
	\lim_{\tau\to \infty} \int_0^\infty \dd t \ed^{-t/\tau} \langle \mathcal J_1 (0) \mathcal J_{1/W} (t) \rangle_W + \mathcal O (\lambda^3)
\end{equation}
where $\langle \cdot \rangle_{W}$ denotes the average over the GGE, and where the dynamics in the time integral is the free dynamics ($\lambda = 0$).
Expression~\eqref{eq: rate p=1} can thus be evaluated explicitly in leading order. 
In the SM, we provide a derivation of eq.~\eqref{eq: rate p=1} which is fully consistent with the derivation used in the more general case $p>1$. 
We tested numerically the validity of eq.~\eqref{eq: rate p=1} for $r=6$ and $\delta = 0.35$, 
for an out-of-equilibrium initial state corresponding to $W = (\beta (\omega (k) - \mu))^{-1}$ with $\mu=-1$ and various values of $\beta$. 
We found excellent agreement with the value of $\gamma(W)$ extracted from direct simulation of the dynamics, 
see Fig.~\ref{fig: r=6 delta=0.35}.
See SM for the numerical protocol.

For $p>1$, we find it convenient to move to the rotated frame where $N_0$ is pseudo-conserved quantity of $\tilde H$, see Claim 1. 
The GGE is now parametrized by only two extensive quantities $\tilde H$ and $N_0$. 
Our main assumption (to be discussed later on) is that the system is in a state $\rho$, $\lambda^p$-close to the GGE $\rho_0$:
\begin{equation}\label{eq: main assumption}
	\rho = \rho_0 \times (1 + \lambda^p f + \mathcal O (\lambda^{p+1})), \quad \rho_0 \sim \ed^{- \beta (\tilde H - \mu N_0 )}
\end{equation}
where $f$ is some correction that will be determined by maximizing local stationarity. 
The key observation is that $\langle J \rangle_{\rho_0} = 0$ where $J := \lambda^p \mathcal J = \lbrace \tilde H , N_0\rbrace $ \cite{mallayya_et_al_2019}. 
Indeed,
$$
	\ed^{- \beta (\tilde H - \mu N_0 )} \lbrace \tilde H , N_0 \rbrace
	= 
	- \frac{1}{\beta}\lbrace \ed^{- \beta (\tilde H - \mu N_0 )} , N_0 \rbrace
$$
and the integral of a Poisson bracket vanishes. 
Hence the instantaneous dissipation rate $\gamma (\beta,\mu)$ may be written as
$$
	\gamma (\beta,\mu) = \langle J \rangle_\rho = \lambda^{2p} \langle \mathcal J f \rangle_{\rho_0} + \mathcal O (\lambda^{2p + 1}). 
$$
Since this rate scales as $\lambda^{2p}$, the state $\rho$ itself evolve on that time scale. 
Stationarity on shorter time scales determines $f$ and an explicit computation yields, see SM:   
\begin{multline}\label{eq: rate p>1}
	\gamma (\beta,\mu) = \frac{\beta \mu\lambda^{2p}}{\delta(0)}	
	\lim_{\tau\to \infty} \int_0^\infty \dd t \ed^{-t/\tau} \langle \mathcal J (0) \mathcal J (t) \rangle_{\rho_0}\\
	 + \mathcal O (\lambda^{2p+1}).
\end{multline}
Again, the dynamics in the time integral is the dynamics generated by $H_0$ and $\gamma (\beta,\mu)$ can thus be computed explicitly in leading order.

\begin{figure}[h]
    \centering
    \includegraphics[draft=false,height = 5cm,width = 7.5cm]{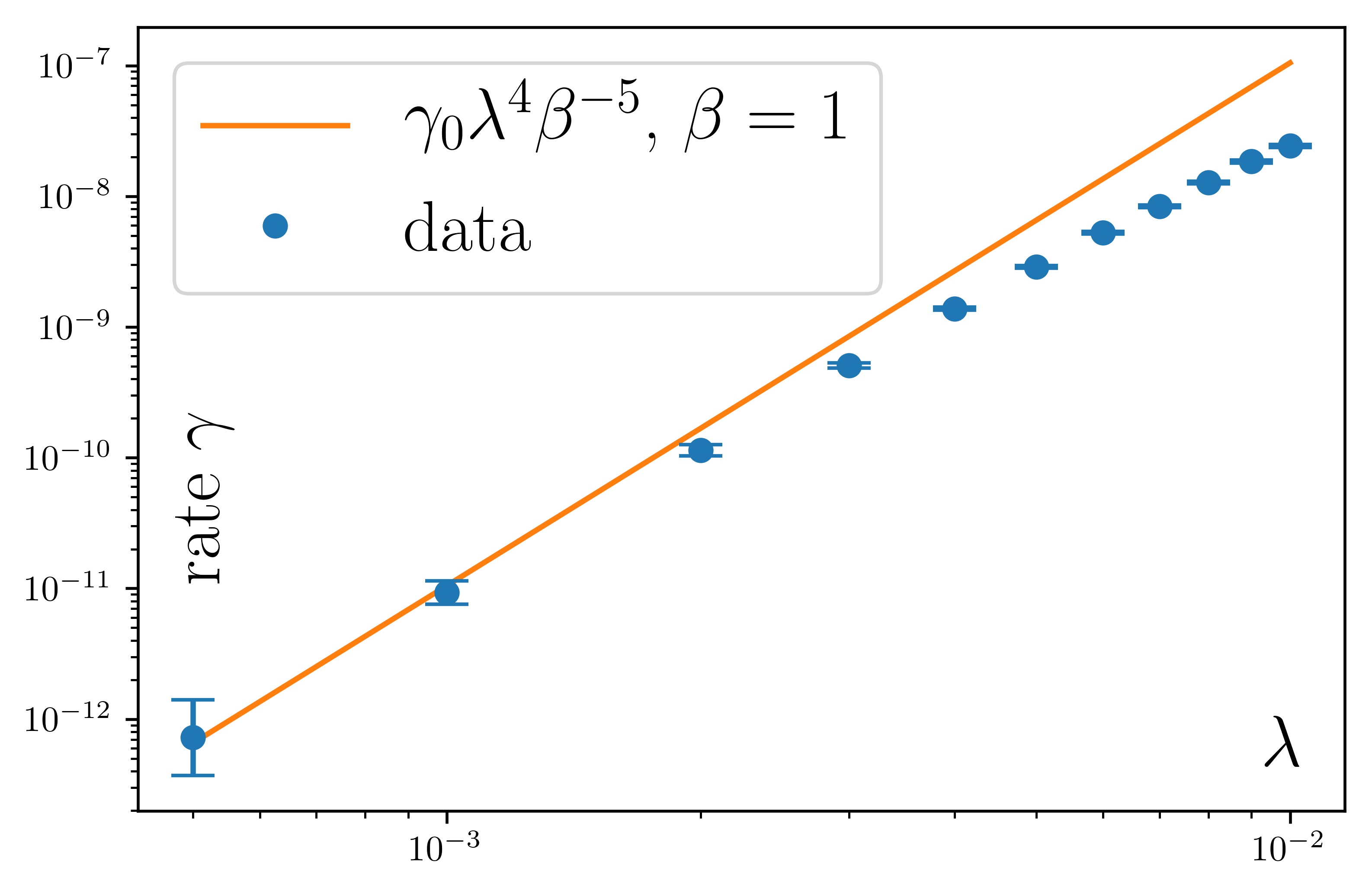}
	\includegraphics[draft=false,height = 5cm,width = 7.5cm]{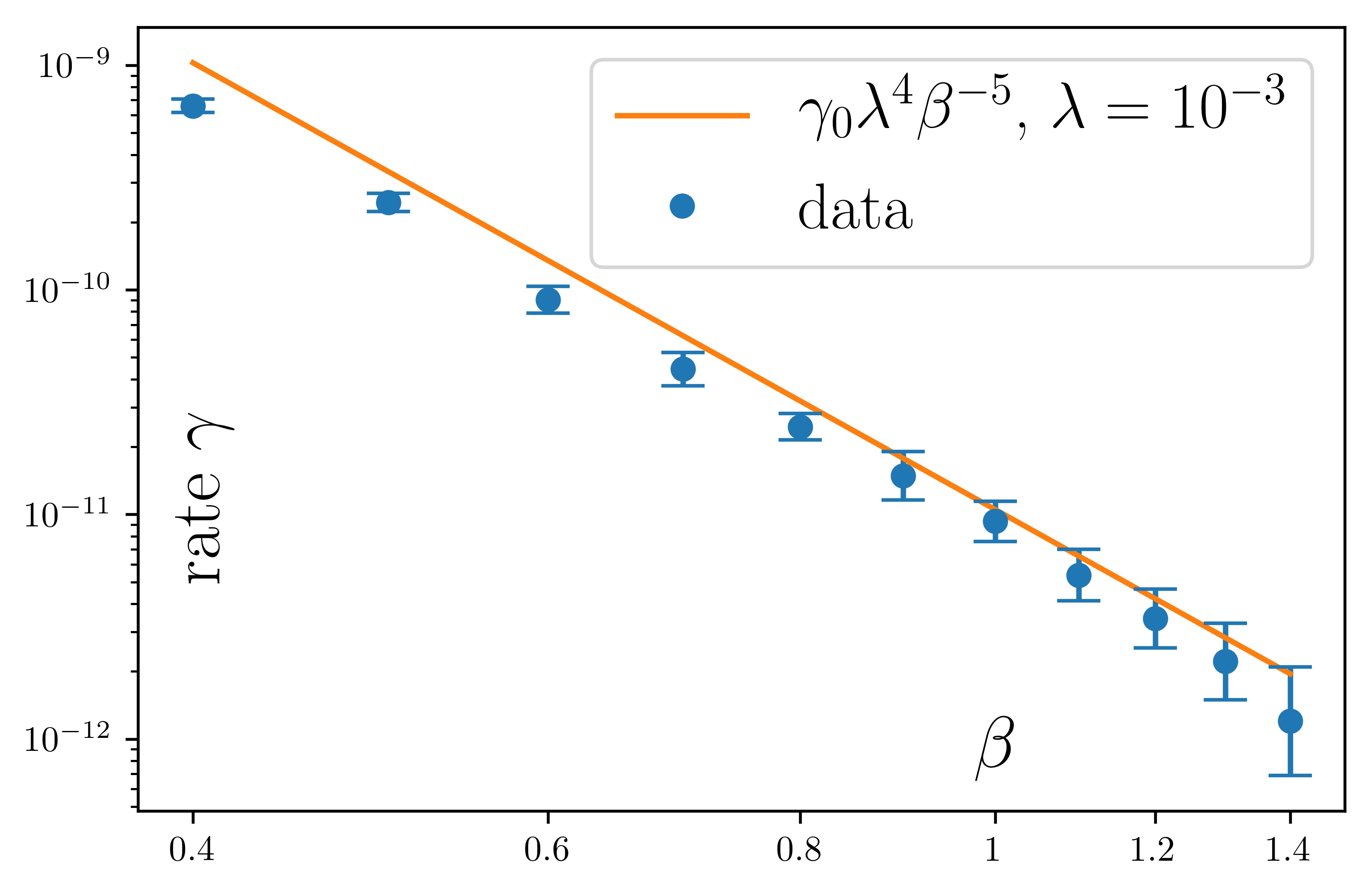}
    \caption{$\gamma$ as a function of $\lambda$ (upper panel) and $\beta$ (lower panel) for $r=4$ and $\delta = 0.45$. 
    Our theory, eq.~\eqref{eq: rate p>1}, predicts $\gamma  = \gamma_0 \lambda^4 \beta^{-5}$ with $\gamma_0\simeq 10.5$.} 
    \label{fig: r=4 delta=0.45}
\end{figure}

\begin{figure}[h]
    \centering
   	\includegraphics[draft=false,height = 5cm,width = 7.5cm]{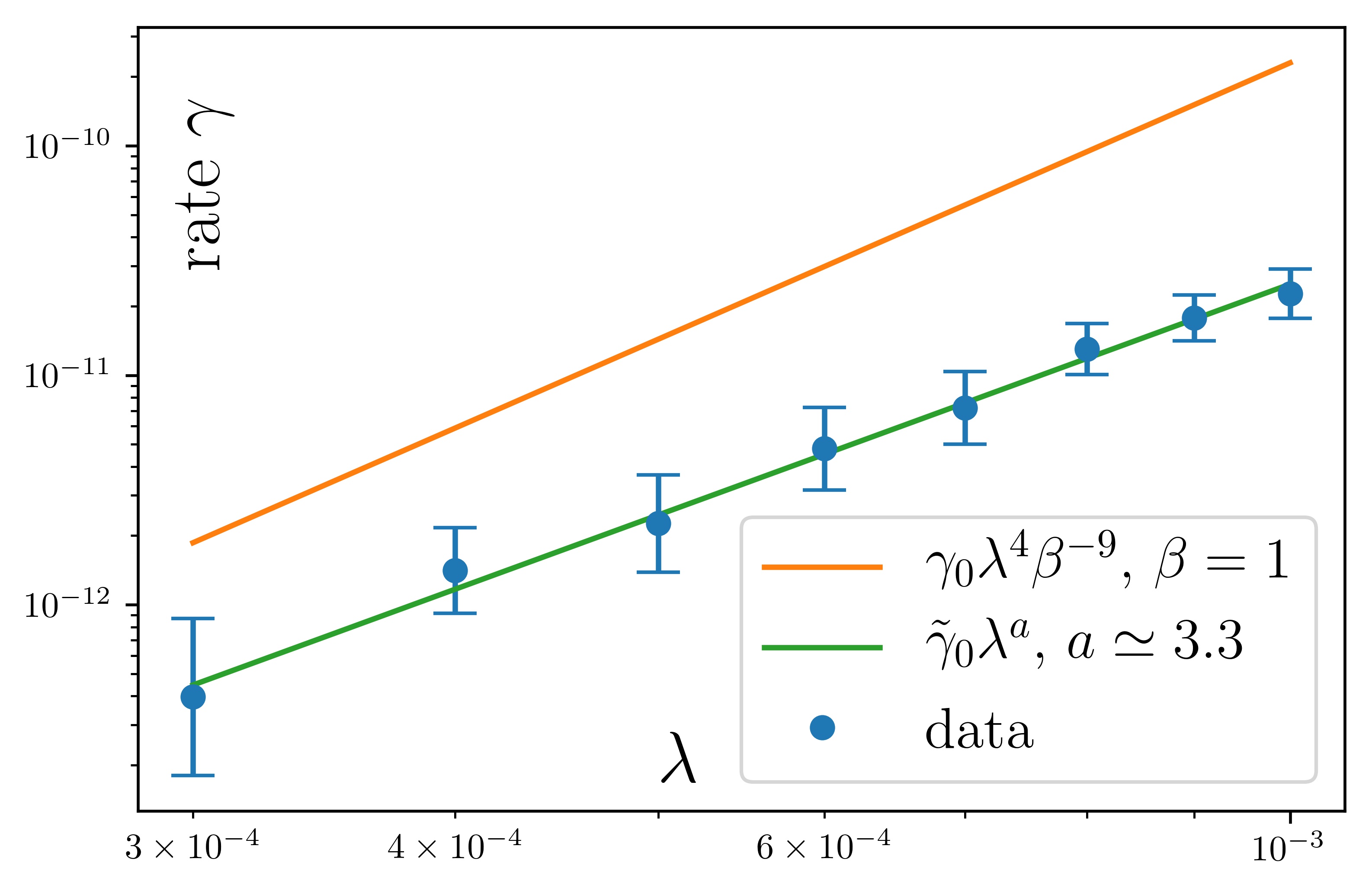}
	\includegraphics[draft=false,height = 5cm,width = 7.5cm]{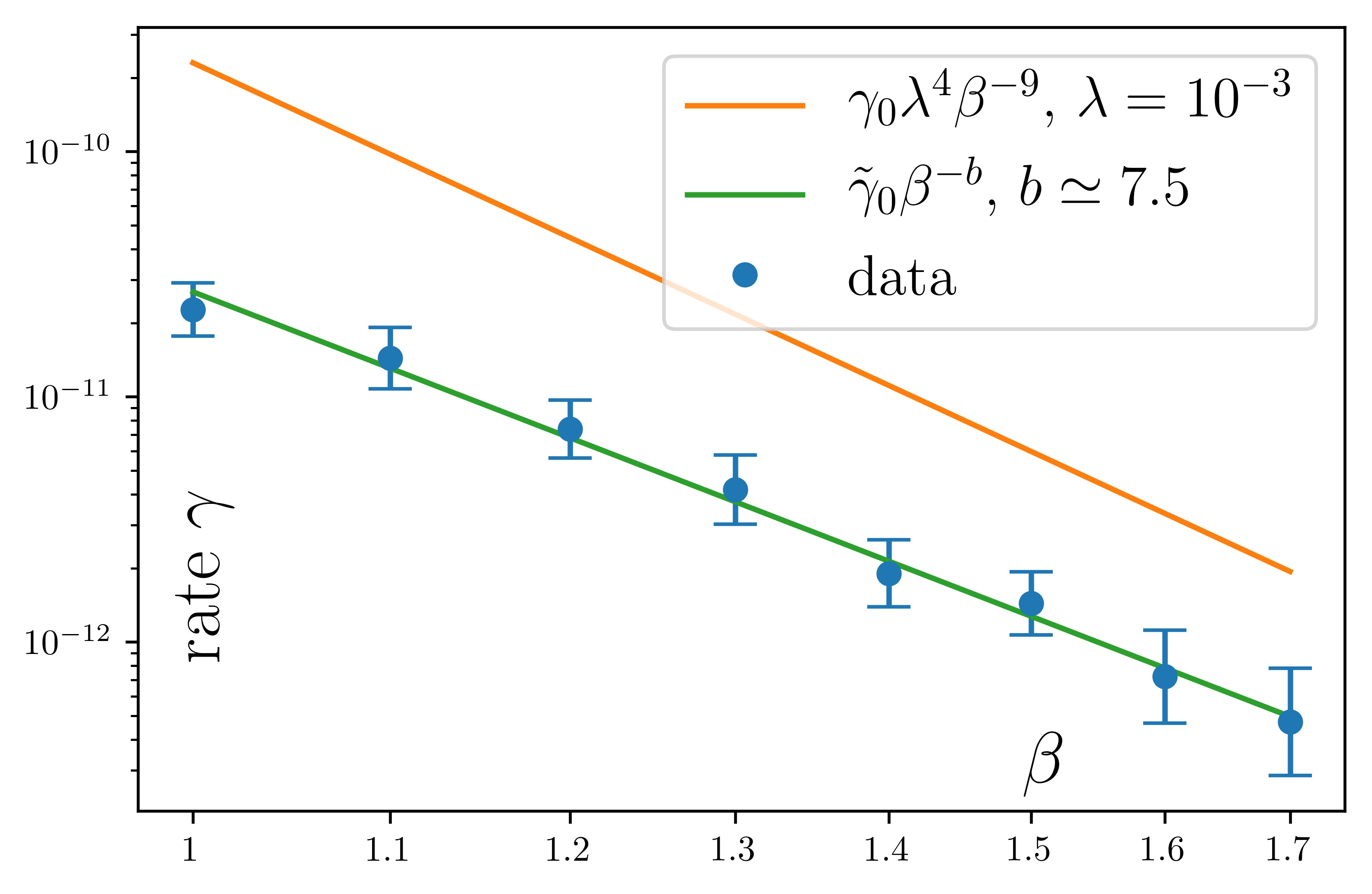}
    \caption{$\gamma$ as a function of $\lambda$ (upper panel) and $\beta$ (lower panel) for $r=6$ and $\delta = 0.28$. 
    Our theory, eq.~\eqref{eq: rate p>1}, predicts $\gamma  = \gamma_0 \lambda^4 \beta^{-9}$ with $\gamma_0\simeq 230$.} 
    \label{fig: r=6 delta=0.28}
\end{figure}

We performed two sets of tests in the case $p=2$
(accessing larger values of $p$ would require too long simulation times).
In all cases, we start from a state of the type $\rho_0$ with $\mu = -1$ and various values of $\beta$.  
For $r=4$ and $\delta=0.45$, the results reported on Fig.~\ref{fig: r=4 delta=0.45} show very good agreement between the prediction from eq.~\eqref{eq: rate p>1}
and direct simulation of the dynamics. 
For $r=6$ and $\delta=0.28$, the observed rate is significantly smaller than the one predicted by eq.~\eqref{eq: rate p>1},
but it decreases slower as a function of $\lambda$ and $\beta^{-1}$, 
see Fig.~\ref{fig: r=6 delta=0.28}. 
Comparing with the discrepancies at the largest values of $\lambda$ on the upper panel of Fig.~\ref{fig: r=4 delta=0.45}, 
makes it plausible that our theory just needs smaller values of $\lambda$ to be validated. 
Besides, the fact that the observations are below the theoretical predictions is a second indication that the theory will become accurate for smaller values of $\lambda$, 
since a smaller rate guarantees that our main hypothesis, eq.~\eqref{eq: main assumption}, from which our predictions follow, is more easily satisfied.

Irrespectively of numerical observations, we finally would like to make a consistency check of the main assumption in eq.~\eqref{eq: main assumption}.
On the one hand, due to the dissipation of $N_0$, the state $\rho_0$ evolves with time at a rate $v_1 \sim \lambda^{2p}$, 
i.e.\@ $\beta,\mu$ evolve at this rate in order to yield correct values for $\langle \tilde H\rangle_{\rho_0}$ and $\langle N_0 \rangle_{\rho_0}$.
On the other hand, the system relaxes towards the instantaneous pseudo-equilibrium $\rho_0$ through kinetic processes.
Assuming that the state $\rho$ is at a distance of order $\lambda^p$ from $\rho_0$, as required by eq.~\eqref{eq: main assumption},
we conclude that it moves at a rate $v_2 \sim \lambda^2 \times \lambda^p$. 
Consistency of the theory requires that $v_1 \lesssim v_2$, 
i.e.\@ that the instantaneous fixed point $\rho_0$ moves slow enough so that the state $\rho$ has the time to  relax to it. 
Clearly, this is wrong for $p=1$, marginal for $p=2$ and fine for $p > 2$. 
We had treated separately the case $p=1$, since indeed there is no reason to think that the pre-thermal state should be characterized by the two parameters $\beta,\mu$ only. 
Unfortunately, the above argument is not conclusive for $p=2$, while numerical data are only available in this case.

\bigskip
\noindent
\textbf{Conclusions and outlook --}
Our work provides a new example of long pre-thermal plateau, 
it shows how a phenomenology initially explored in quantum systems carries over to a classical set-up, 
and it participates to recent efforts to describe accurately the dissipation of pseudo-conserved quantities. 
The main features of our theory carry over to $d>1$ and, for $d=3$, 
we may contemplate the possibility of realizing a pre-thermal Bose-Einstein condensate in this classical system, 
exploiting the conservation of the number of phonons over a very long period.

\bigskip

\begin{acknowledgments} 
	We thank W.~De Roeck and H.~Spohn for helpful discussions, and C.~Mendl for providing the original code for numerical simulations. 
	F.~H.\@ and J.~L.\@ benefited from the support of the project EDNHS ANR-14-CE25-0011, 
	and F.~H.\@ from the project LSD ANR-15-CE40-0020-01 of the French National Research Agency (ANR), 
	as well as from the support of the International Centre for Theoretical Sciences (ICTS) during a visit for the program 
	- Thermalization, Many body localization and Hydrodynamics (Code: ICTS/hydrodynamics2019/11).
	The work has also been supported by the Academy of Finland via the Centre of Excellence in Analysis and Dynamics Research (project 307333) 
	and the Matter and Materials Profi4 university profiling action.
\end{acknowledgments}

\bibliography{pre_thermal_bibliography}

\begin{thebibliography}{39}%
\makeatletter
\providecommand \@ifxundefined [1]{%
 \@ifx{#1\undefined}
}%
\providecommand \@ifnum [1]{%
 \ifnum #1\expandafter \@firstoftwo
 \else \expandafter \@secondoftwo
 \fi
}%
\providecommand \@ifx [1]{%
 \ifx #1\expandafter \@firstoftwo
 \else \expandafter \@secondoftwo
 \fi
}%
\providecommand \natexlab [1]{#1}%
\providecommand \enquote  [1]{``#1''}%
\providecommand \bibnamefont  [1]{#1}%
\providecommand \bibfnamefont [1]{#1}%
\providecommand \citenamefont [1]{#1}%
\providecommand \href@noop [0]{\@secondoftwo}%
\providecommand \href [0]{\begingroup \@sanitize@url \@href}%
\providecommand \@href[1]{\@@startlink{#1}\@@href}%
\providecommand \@@href[1]{\endgroup#1\@@endlink}%
\providecommand \@sanitize@url [0]{\catcode `\\12\catcode `\$12\catcode
  `\&12\catcode `\#12\catcode `\^12\catcode `\_12\catcode `\%12\relax}%
\providecommand \@@startlink[1]{}%
\providecommand \@@endlink[0]{}%
\providecommand \url  [0]{\begingroup\@sanitize@url \@url }%
\providecommand \@url [1]{\endgroup\@href {#1}{\urlprefix }}%
\providecommand \urlprefix  [0]{URL }%
\providecommand \Eprint [0]{\href }%
\providecommand \doibase [0]{http://dx.doi.org/}%
\providecommand \selectlanguage [0]{\@gobble}%
\providecommand \bibinfo  [0]{\@secondoftwo}%
\providecommand \bibfield  [0]{\@secondoftwo}%
\providecommand \translation [1]{[#1]}%
\providecommand \BibitemOpen [0]{}%
\providecommand \bibitemStop [0]{}%
\providecommand \bibitemNoStop [0]{.\EOS\space}%
\providecommand \EOS [0]{\spacefactor3000\relax}%
\providecommand \BibitemShut  [1]{\csname bibitem#1\endcsname}%
\let\auto@bib@innerbib\@empty
\bibitem [{\citenamefont {Anderson}(1958)}]{anderson_1958}%
  \BibitemOpen
  \bibfield  {author} {\bibinfo {author} {\bibfnamefont {P.~W.}\ \bibnamefont
  {Anderson}},\ }\bibfield  {title} {\enquote {\bibinfo {title} {Absence of
  diffusion in certain random lattices},}\ }\href {\doibase
  10.1103/PhysRev.109.1492} {\bibfield  {journal} {\bibinfo  {journal}
  {Physical Review}\ }\textbf {\bibinfo {volume} {109}},\ \bibinfo {pages}
  {1492--1505} (\bibinfo {year} {1958})}\BibitemShut {NoStop}%
\bibitem [{\citenamefont {Gornyi}\ \emph {et~al.}(2005)\citenamefont {Gornyi},
  \citenamefont {Mirlin},\ and\ \citenamefont {Polyakov}}]{gornyi_et_al_2005}%
  \BibitemOpen
  \bibfield  {author} {\bibinfo {author} {\bibfnamefont {I.}~\bibnamefont
  {Gornyi}}, \bibinfo {author} {\bibfnamefont {A.}~\bibnamefont {Mirlin}}, \
  and\ \bibinfo {author} {\bibfnamefont {D.}~\bibnamefont {Polyakov}},\
  }\bibfield  {title} {\enquote {\bibinfo {title} {{Interacting electrons in
  disordered wires: Anderson localization and low-T transport}},}\ }\href@noop
  {} {\bibfield  {journal} {\bibinfo  {journal} {Physical Review Letters}\
  }\textbf {\bibinfo {volume} {95}},\ \bibinfo {pages} {206603} (\bibinfo
  {year} {2005})}\BibitemShut {NoStop}%
\bibitem [{\citenamefont {Basko}\ \emph {et~al.}(2006)\citenamefont {Basko},
  \citenamefont {Aleiner},\ and\ \citenamefont {Altshuler}}]{basko_et_al_2006}%
  \BibitemOpen
  \bibfield  {author} {\bibinfo {author} {\bibfnamefont {D.~M.}\ \bibnamefont
  {Basko}}, \bibinfo {author} {\bibfnamefont {I.~L.}\ \bibnamefont {Aleiner}},
  \ and\ \bibinfo {author} {\bibfnamefont {B.~L.}\ \bibnamefont {Altshuler}},\
  }\bibfield  {title} {\enquote {\bibinfo {title} {{Metal–insulator
  transition in a weakly interacting many-electron system with localized
  single-particle states}},}\ }\href {\doibase
  https://doi.org/10.1016/j.aop.2005.11.014} {\bibfield  {journal} {\bibinfo
  {journal} {Annals of Physics}\ }\textbf {\bibinfo {volume} {321}},\ \bibinfo
  {pages} {1126--1205} (\bibinfo {year} {2006})}\BibitemShut {NoStop}%
\bibitem [{\citenamefont {Turner}\ \emph {et~al.}(2018)\citenamefont {Turner},
  \citenamefont {Michailidis}, \citenamefont {Abanin}, \citenamefont {Serbyn},\
  and\ \citenamefont {Papic}}]{turner_et_al_18}%
  \BibitemOpen
  \bibfield  {author} {\bibinfo {author} {\bibfnamefont {C.~J.}\ \bibnamefont
  {Turner}}, \bibinfo {author} {\bibfnamefont {A.~A.}\ \bibnamefont
  {Michailidis}}, \bibinfo {author} {\bibfnamefont {D.~A.}\ \bibnamefont
  {Abanin}}, \bibinfo {author} {\bibfnamefont {M.}~\bibnamefont {Serbyn}}, \
  and\ \bibinfo {author} {\bibfnamefont {Z.}~\bibnamefont {Papic}},\ }\bibfield
   {title} {\enquote {\bibinfo {title} {{Weak ergodicity breaking from quantum
  many-body scars}},}\ }\href {\doibase 10.1038/s41567-018-0137-5} {\bibfield
  {journal} {\bibinfo  {journal} {Nature Physics}\ }\textbf {\bibinfo {volume}
  {14}},\ \bibinfo {pages} {745--749} (\bibinfo {year} {2018})}\BibitemShut
  {NoStop}%
\bibitem [{\citenamefont {Kagan}\ and\ \citenamefont
  {Maksimov}(1984)}]{kagan_maksimov_1984}%
  \BibitemOpen
  \bibfield  {author} {\bibinfo {author} {\bibfnamefont {Y.}~\bibnamefont
  {Kagan}}\ and\ \bibinfo {author} {\bibfnamefont {L.}~\bibnamefont
  {Maksimov}},\ }\bibfield  {title} {\enquote {\bibinfo {title} {{Localization
  in a system of interacting particles diffusing in a regular crystal}},}\
  }\href@noop {} {\bibfield  {journal} {\bibinfo  {journal} {Zhurnal
  Eksperimental’noi i Teoreticheskoi Fiziki}\ }\textbf {\bibinfo {volume}
  {87}},\ \bibinfo {pages} {348--365} (\bibinfo {year} {1984})}\BibitemShut
  {NoStop}%
\bibitem [{\citenamefont {Fermi}\ \emph {et~al.}(1955)\citenamefont {Fermi},
  \citenamefont {Pasta}, \citenamefont {Ulam},\ and\ \citenamefont
  {Tsingou}}]{fermi_et_al_1955}%
  \BibitemOpen
  \bibfield  {author} {\bibinfo {author} {\bibfnamefont {E.}~\bibnamefont
  {Fermi}}, \bibinfo {author} {\bibfnamefont {P.}~\bibnamefont {Pasta}},
  \bibinfo {author} {\bibfnamefont {S.}~\bibnamefont {Ulam}}, \ and\ \bibinfo
  {author} {\bibfnamefont {M.}~\bibnamefont {Tsingou}},\ }\href@noop {} {\emph
  {\bibinfo {title} {{Studies of the nonlinear problems}}}},\ \bibinfo {type}
  {Tech. Rep.}\ (\bibinfo  {institution} {Los Alamos Scientific Lab., N.
  Mex.},\ \bibinfo {year} {1955})\BibitemShut {NoStop}%
\bibitem [{\citenamefont {Basko}(2011)}]{basko_2011}%
  \BibitemOpen
  \bibfield  {author} {\bibinfo {author} {\bibfnamefont {D.}~\bibnamefont
  {Basko}},\ }\bibfield  {title} {\enquote {\bibinfo {title} {{Weak chaos in
  the disordered nonlinear Schr{\"o}dinger chain: destruction of Anderson
  localization by Arnold diffusion}},}\ }\href@noop {} {\bibfield  {journal}
  {\bibinfo  {journal} {Annals of Physics}\ }\textbf {\bibinfo {volume}
  {326}},\ \bibinfo {pages} {1577--1655} (\bibinfo {year} {2011})}\BibitemShut
  {NoStop}%
\bibitem [{\citenamefont {D’Alessio}\ and\ \citenamefont
  {Polkovnikov}(2013)}]{dalessio_polkovnikov_2013}%
  \BibitemOpen
  \bibfield  {author} {\bibinfo {author} {\bibfnamefont {L.}~\bibnamefont
  {D’Alessio}}\ and\ \bibinfo {author} {\bibfnamefont {A.}~\bibnamefont
  {Polkovnikov}},\ }\bibfield  {title} {\enquote {\bibinfo {title} {{Many-body
  energy localization transition in periodically driven systems}},}\
  }\href@noop {} {\bibfield  {journal} {\bibinfo  {journal} {Annals of
  Physics}\ }\textbf {\bibinfo {volume} {333}},\ \bibinfo {pages} {19--33}
  (\bibinfo {year} {2013})}\BibitemShut {NoStop}%
\bibitem [{\citenamefont {D'Alessio}\ and\ \citenamefont
  {Rigol}(2014)}]{dalessio_rigol_2014}%
  \BibitemOpen
  \bibfield  {author} {\bibinfo {author} {\bibfnamefont {L.}~\bibnamefont
  {D'Alessio}}\ and\ \bibinfo {author} {\bibfnamefont {M.}~\bibnamefont
  {Rigol}},\ }\bibfield  {title} {\enquote {\bibinfo {title} {{Long-time
  Behavior of Isolated Periodically Driven Interacting Lattice Systems}},}\
  }\href {\doibase 10.1103/PhysRevX.4.041048} {\bibfield  {journal} {\bibinfo
  {journal} {Physical Review X}\ }\textbf {\bibinfo {volume} {4}},\ \bibinfo
  {pages} {041048} (\bibinfo {year} {2014})}\BibitemShut {NoStop}%
\bibitem [{\citenamefont {Lazarides}\ \emph {et~al.}(2014)\citenamefont
  {Lazarides}, \citenamefont {Das},\ and\ \citenamefont
  {Moessner}}]{lazarides_et_al_2014}%
  \BibitemOpen
  \bibfield  {author} {\bibinfo {author} {\bibfnamefont {A.}~\bibnamefont
  {Lazarides}}, \bibinfo {author} {\bibfnamefont {A.}~\bibnamefont {Das}}, \
  and\ \bibinfo {author} {\bibfnamefont {R.}~\bibnamefont {Moessner}},\
  }\bibfield  {title} {\enquote {\bibinfo {title} {{Equilibrium states of
  generic quantum systems subject to periodic driving}},}\ }\href {\doibase
  10.1103/PhysRevE.90.012110} {\bibfield  {journal} {\bibinfo  {journal}
  {Physical Review E}\ }\textbf {\bibinfo {volume} {90}},\ \bibinfo {pages}
  {012110} (\bibinfo {year} {2014})}\BibitemShut {NoStop}%
\bibitem [{\citenamefont {Abanin}\ \emph {et~al.}(2015)\citenamefont {Abanin},
  \citenamefont {De~Roeck},\ and\ \citenamefont
  {Huveneers}}]{abanin_et_al_2015}%
  \BibitemOpen
  \bibfield  {author} {\bibinfo {author} {\bibfnamefont {D.~A.}\ \bibnamefont
  {Abanin}}, \bibinfo {author} {\bibfnamefont {W.}~\bibnamefont {De~Roeck}}, \
  and\ \bibinfo {author} {\bibfnamefont {F.}~\bibnamefont {Huveneers}},\
  }\bibfield  {title} {\enquote {\bibinfo {title} {{Exponentially Slow Heating
  in Periodically Driven Many-Body Systems}},}\ }\href {\doibase
  10.1103/PhysRevLett.115.256803} {\bibfield  {journal} {\bibinfo  {journal}
  {Physical Review Letters}\ }\textbf {\bibinfo {volume} {115}},\ \bibinfo
  {pages} {256803} (\bibinfo {year} {2015})}\BibitemShut {NoStop}%
\bibitem [{\citenamefont {Abanin}\ \emph
  {et~al.}(2017{\natexlab{a}})\citenamefont {Abanin}, \citenamefont {De~Roeck},
  \citenamefont {Ho},\ and\ \citenamefont {Huveneers}}]{abanin_et_al_prb_2017}%
  \BibitemOpen
  \bibfield  {author} {\bibinfo {author} {\bibfnamefont {D.~A.}\ \bibnamefont
  {Abanin}}, \bibinfo {author} {\bibfnamefont {W.}~\bibnamefont {De~Roeck}},
  \bibinfo {author} {\bibfnamefont {W.~W.}\ \bibnamefont {Ho}}, \ and\ \bibinfo
  {author} {\bibfnamefont {F.}~\bibnamefont {Huveneers}},\ }\bibfield  {title}
  {\enquote {\bibinfo {title} {{Effective Hamiltonians, prethermalization, and
  slow energy absorption in periodically driven many-body systems}},}\ }\href
  {\doibase 10.1103/PhysRevB.95.014112} {\bibfield  {journal} {\bibinfo
  {journal} {Physical Review B}\ }\textbf {\bibinfo {volume} {95}},\ \bibinfo
  {pages} {014112} (\bibinfo {year} {2017}{\natexlab{a}})}\BibitemShut
  {NoStop}%
\bibitem [{\citenamefont {Abanin}\ \emph
  {et~al.}(2017{\natexlab{b}})\citenamefont {Abanin}, \citenamefont {De~Roeck},
  \citenamefont {Ho},\ and\ \citenamefont {Huveneers}}]{abanin_et_al_cmp_2017}%
  \BibitemOpen
  \bibfield  {author} {\bibinfo {author} {\bibfnamefont {D.~A.}\ \bibnamefont
  {Abanin}}, \bibinfo {author} {\bibfnamefont {W.}~\bibnamefont {De~Roeck}},
  \bibinfo {author} {\bibfnamefont {W.~W.}\ \bibnamefont {Ho}}, \ and\ \bibinfo
  {author} {\bibfnamefont {F.}~\bibnamefont {Huveneers}},\ }\bibfield  {title}
  {\enquote {\bibinfo {title} {{A Rigorous Theory of Many-Body
  Prethermalization for Periodically Driven and Closed Quantum Systems}},}\
  }\href {\doibase 10.1007/s00220-017-2930-x} {\bibfield  {journal} {\bibinfo
  {journal} {Communications in Mathematical Physics}\ }\textbf {\bibinfo
  {volume} {354}},\ \bibinfo {pages} {809--827} (\bibinfo {year}
  {2017}{\natexlab{b}})}\BibitemShut {NoStop}%
\bibitem [{\citenamefont {Mori}\ \emph {et~al.}(2016)\citenamefont {Mori},
  \citenamefont {Kuwahara},\ and\ \citenamefont {Saito}}]{mori_et_al_2016}%
  \BibitemOpen
  \bibfield  {author} {\bibinfo {author} {\bibfnamefont {T.}~\bibnamefont
  {Mori}}, \bibinfo {author} {\bibfnamefont {T.}~\bibnamefont {Kuwahara}}, \
  and\ \bibinfo {author} {\bibfnamefont {K.}~\bibnamefont {Saito}},\ }\bibfield
   {title} {\enquote {\bibinfo {title} {{Rigorous Bound on Energy Absorption
  and Generic Relaxation in Periodically Driven Quantum Systems}},}\ }\href
  {\doibase 10.1103/PhysRevLett.116.120401} {\bibfield  {journal} {\bibinfo
  {journal} {Physical Review Letters}\ }\textbf {\bibinfo {volume} {116}},\
  \bibinfo {pages} {120401} (\bibinfo {year} {2016})}\BibitemShut {NoStop}%
\bibitem [{\citenamefont {Sensarma}\ \emph {et~al.}(2010)\citenamefont
  {Sensarma}, \citenamefont {Pekker}, \citenamefont {Altman}, \citenamefont
  {Demler}, \citenamefont {Strohmaier}, \citenamefont {Greif}, \citenamefont
  {J{\"o}rdens}, \citenamefont {Tarruell}, \citenamefont {Moritz},\ and\
  \citenamefont {Esslinger}}]{sensarma_et_al_2010}%
  \BibitemOpen
  \bibfield  {author} {\bibinfo {author} {\bibfnamefont {R.}~\bibnamefont
  {Sensarma}}, \bibinfo {author} {\bibfnamefont {D.}~\bibnamefont {Pekker}},
  \bibinfo {author} {\bibfnamefont {E.}~\bibnamefont {Altman}}, \bibinfo
  {author} {\bibfnamefont {E.}~\bibnamefont {Demler}}, \bibinfo {author}
  {\bibfnamefont {N.}~\bibnamefont {Strohmaier}}, \bibinfo {author}
  {\bibfnamefont {D.}~\bibnamefont {Greif}}, \bibinfo {author} {\bibfnamefont
  {R.}~\bibnamefont {J{\"o}rdens}}, \bibinfo {author} {\bibfnamefont
  {L.}~\bibnamefont {Tarruell}}, \bibinfo {author} {\bibfnamefont
  {H.}~\bibnamefont {Moritz}}, \ and\ \bibinfo {author} {\bibfnamefont
  {T.}~\bibnamefont {Esslinger}},\ }\bibfield  {title} {\enquote {\bibinfo
  {title} {{Lifetime of double occupancies in the Fermi-Hubbard model}},}\
  }\href {\doibase 10.1103/PhysRevB.82.224302} {\bibfield  {journal} {\bibinfo
  {journal} {Physical Review B}\ }\textbf {\bibinfo {volume} {82}},\ \bibinfo
  {pages} {224302} (\bibinfo {year} {2010})}\BibitemShut {NoStop}%
\bibitem [{\citenamefont {Vajna}\ \emph {et~al.}(2018)\citenamefont {Vajna},
  \citenamefont {Klobas}, \citenamefont {Prosen},\ and\ \citenamefont
  {Polkovnikov}}]{vajna_et_al_2018}%
  \BibitemOpen
  \bibfield  {author} {\bibinfo {author} {\bibfnamefont {S.}~\bibnamefont
  {Vajna}}, \bibinfo {author} {\bibfnamefont {K.}~\bibnamefont {Klobas}},
  \bibinfo {author} {\bibfnamefont {T.}~\bibnamefont {Prosen}}, \ and\ \bibinfo
  {author} {\bibfnamefont {A.}~\bibnamefont {Polkovnikov}},\ }\bibfield
  {title} {\enquote {\bibinfo {title} {{Replica Resummation of the
  Baker-Campbell-Hausdorff Series}},}\ }\href {\doibase
  10.1103/PhysRevLett.120.200607} {\bibfield  {journal} {\bibinfo  {journal}
  {Physical Review Letters}\ }\textbf {\bibinfo {volume} {120}},\ \bibinfo
  {pages} {200607} (\bibinfo {year} {2018})}\BibitemShut {NoStop}%
\bibitem [{\citenamefont {De~Roeck}\ and\ \citenamefont
  {Verreet}(2019)}]{de_roeck_verreet_2019}%
  \BibitemOpen
  \bibfield  {author} {\bibinfo {author} {\bibfnamefont {W.}~\bibnamefont
  {De~Roeck}}\ and\ \bibinfo {author} {\bibfnamefont {V.}~\bibnamefont
  {Verreet}},\ }\bibfield  {title} {\enquote {\bibinfo {title} {{Very slow
  heating for weakly driven quantum many-body systems}},}\ }\href@noop {}
  {\bibfield  {journal} {\bibinfo  {journal} {arXiv}\ } (\bibinfo {year}
  {2019})},\ \Eprint {http://arxiv.org/abs/1911.01998} {arXiv:1911.01998
  [cond-mat.stat-mech]} \BibitemShut {NoStop}%
\bibitem [{\citenamefont {Carati}\ and\ \citenamefont
  {Maiocchi}(2012)}]{carati_maiocchi_2012}%
  \BibitemOpen
  \bibfield  {author} {\bibinfo {author} {\bibfnamefont {A.}~\bibnamefont
  {Carati}}\ and\ \bibinfo {author} {\bibfnamefont {A.~M.}\ \bibnamefont
  {Maiocchi}},\ }\bibfield  {title} {\enquote {\bibinfo {title} {{Exponentially
  Long Stability Times for a Nonlinear Lattice in the Thermodynamic Limit}},}\
  }\href {https://doi.org/10.1007/s00220-012-1522-z} {\bibfield  {journal}
  {\bibinfo  {journal} {Communications in Mathematical Physics}\ }\textbf
  {\bibinfo {volume} {314}},\ \bibinfo {pages} {129--161} (\bibinfo {year}
  {2012})}\BibitemShut {NoStop}%
\bibitem [{\citenamefont {Giorgilli}\ \emph {et~al.}(2015)\citenamefont
  {Giorgilli}, \citenamefont {Paleari},\ and\ \citenamefont
  {Penati}}]{giorgilli_et_al_2015}%
  \BibitemOpen
  \bibfield  {author} {\bibinfo {author} {\bibfnamefont {A.}~\bibnamefont
  {Giorgilli}}, \bibinfo {author} {\bibfnamefont {S.}~\bibnamefont {Paleari}},
  \ and\ \bibinfo {author} {\bibfnamefont {T.}~\bibnamefont {Penati}},\
  }\bibfield  {title} {\enquote {\bibinfo {title} {{An extensive adiabatic
  invariant for the Klein--Gordon model in the thermodynamic limit}},}\
  }\href@noop {} {\bibfield  {journal} {\bibinfo  {journal} {Annales Henri
  Poincar{\'e}}\ }\textbf {\bibinfo {volume} {16}},\ \bibinfo {pages}
  {897--959} (\bibinfo {year} {2015})}\BibitemShut {NoStop}%
\bibitem [{\citenamefont {Howell}\ \emph {et~al.}(2019)\citenamefont {Howell},
  \citenamefont {Weinberg}, \citenamefont {Sels}, \citenamefont {Polkovnikov},\
  and\ \citenamefont {Bukov}}]{howell_et_al_2019}%
  \BibitemOpen
  \bibfield  {author} {\bibinfo {author} {\bibfnamefont {O.}~\bibnamefont
  {Howell}}, \bibinfo {author} {\bibfnamefont {P.}~\bibnamefont {Weinberg}},
  \bibinfo {author} {\bibfnamefont {D.}~\bibnamefont {Sels}}, \bibinfo {author}
  {\bibfnamefont {A.}~\bibnamefont {Polkovnikov}}, \ and\ \bibinfo {author}
  {\bibfnamefont {M.}~\bibnamefont {Bukov}},\ }\bibfield  {title} {\enquote
  {\bibinfo {title} {{Asymptotic Prethermalization in Periodically Driven
  Classical Spin Chains}},}\ }\href {\doibase 10.1103/PhysRevLett.122.010602}
  {\bibfield  {journal} {\bibinfo  {journal} {Physical Review Letters}\
  }\textbf {\bibinfo {volume} {122}},\ \bibinfo {pages} {010602} (\bibinfo
  {year} {2019})}\BibitemShut {NoStop}%
\bibitem [{\citenamefont {Else}\ \emph
  {et~al.}(2017{\natexlab{a}})\citenamefont {Else}, \citenamefont {Bauer},\
  and\ \citenamefont {Nayak}}]{else_et_al_2017}%
  \BibitemOpen
  \bibfield  {author} {\bibinfo {author} {\bibfnamefont {D.~V.}\ \bibnamefont
  {Else}}, \bibinfo {author} {\bibfnamefont {B.}~\bibnamefont {Bauer}}, \ and\
  \bibinfo {author} {\bibfnamefont {C.}~\bibnamefont {Nayak}},\ }\bibfield
  {title} {\enquote {\bibinfo {title} {{Prethermal Phases of Matter Protected
  by Time-Translation Symmetry}},}\ }\href {\doibase 10.1103/PhysRevX.7.011026}
  {\bibfield  {journal} {\bibinfo  {journal} {Physical Review X}\ }\textbf
  {\bibinfo {volume} {7}},\ \bibinfo {pages} {011026} (\bibinfo {year}
  {2017}{\natexlab{a}})}\BibitemShut {NoStop}%
\bibitem [{\citenamefont {Else}\ \emph {et~al.}(2019)\citenamefont {Else},
  \citenamefont {Ho},\ and\ \citenamefont {Dumitrescu}}]{else_et_al_2019}%
  \BibitemOpen
  \bibfield  {author} {\bibinfo {author} {\bibfnamefont {D.~V.}\ \bibnamefont
  {Else}}, \bibinfo {author} {\bibfnamefont {W.~W}\ \bibnamefont {Ho}}, \ and\
  \bibinfo {author} {\bibfnamefont {P.~T.}\ \bibnamefont {Dumitrescu}},\
  }\bibfield  {title} {\enquote {\bibinfo {title} {{Long-lived interacting
  phases of matter protected by multiple time-translation symmetries in
  quasiperiodically-driven systems}},}\ }\href@noop {} {\bibfield  {journal}
  {\bibinfo  {journal} {arXiv}\ } (\bibinfo {year} {2019})},\ \Eprint
  {http://arxiv.org/abs/1910.03584} {arXiv:1910.03584 [cond-mat.str-el]}
  \BibitemShut {NoStop}%
\bibitem [{\citenamefont {Else}\ \emph
  {et~al.}(2017{\natexlab{b}})\citenamefont {Else}, \citenamefont {Fendley},
  \citenamefont {Kemp},\ and\ \citenamefont {Nayak}}]{else_fendley_et_al_2017}%
  \BibitemOpen
  \bibfield  {author} {\bibinfo {author} {\bibfnamefont {D.~V.}\ \bibnamefont
  {Else}}, \bibinfo {author} {\bibfnamefont {P.}~\bibnamefont {Fendley}},
  \bibinfo {author} {\bibfnamefont {J.}~\bibnamefont {Kemp}}, \ and\ \bibinfo
  {author} {\bibfnamefont {C.}~\bibnamefont {Nayak}},\ }\bibfield  {title}
  {\enquote {\bibinfo {title} {{Prethermal Strong Zero Modes and Topological
  Qubits}},}\ }\href {\doibase 10.1103/PhysRevX.7.041062} {\bibfield  {journal}
  {\bibinfo  {journal} {Physical Review X}\ }\textbf {\bibinfo {volume} {7}},\
  \bibinfo {pages} {041062} (\bibinfo {year} {2017}{\natexlab{b}})}\BibitemShut
  {NoStop}%
\bibitem [{\citenamefont {Lindner}\ \emph {et~al.}(2017)\citenamefont
  {Lindner}, \citenamefont {Berg},\ and\ \citenamefont
  {Rudner}}]{lindner_et_al_2017}%
  \BibitemOpen
  \bibfield  {author} {\bibinfo {author} {\bibfnamefont {N.~H.}\ \bibnamefont
  {Lindner}}, \bibinfo {author} {\bibfnamefont {E.}~\bibnamefont {Berg}}, \
  and\ \bibinfo {author} {\bibfnamefont {M.~S.}\ \bibnamefont {Rudner}},\
  }\bibfield  {title} {\enquote {\bibinfo {title} {Universal chiral quasisteady
  states in periodically driven many-body systems},}\ }\href {\doibase
  10.1103/PhysRevX.7.011018} {\bibfield  {journal} {\bibinfo  {journal}
  {Physical Review X}\ }\textbf {\bibinfo {volume} {7}},\ \bibinfo {pages}
  {011018} (\bibinfo {year} {2017})}\BibitemShut {NoStop}%
\bibitem [{\citenamefont {Martin}\ \emph {et~al.}(2017)\citenamefont {Martin},
  \citenamefont {Refael},\ and\ \citenamefont {Halperin}}]{martin_et_al_2017}%
  \BibitemOpen
  \bibfield  {author} {\bibinfo {author} {\bibfnamefont {I.}~\bibnamefont
  {Martin}}, \bibinfo {author} {\bibfnamefont {G.}~\bibnamefont {Refael}}, \
  and\ \bibinfo {author} {\bibfnamefont {B.}~\bibnamefont {Halperin}},\
  }\bibfield  {title} {\enquote {\bibinfo {title} {{Topological Frequency
  Conversion in Strongly Driven Quantum Systems}},}\ }\href {\doibase
  10.1103/PhysRevX.7.041008} {\bibfield  {journal} {\bibinfo  {journal}
  {Physical Review X}\ }\textbf {\bibinfo {volume} {7}},\ \bibinfo {pages}
  {041008} (\bibinfo {year} {2017})}\BibitemShut {NoStop}%
\bibitem [{\citenamefont {{Dudnikova, T.~V. and Komech, A.~I. and Spohn,
  H.}}(2003)}]{dudnikova_et_al_2003}%
  \BibitemOpen
  \bibfield  {author} {\bibinfo {author} {\bibnamefont {{Dudnikova, T.~V. and
  Komech, A.~I. and Spohn, H.}}},\ }\bibfield  {title} {\enquote {\bibinfo
  {title} {{On the convergence to statistical equilibrium for harmonic
  crystals}},}\ }\href {\doibase 10.1063/1.1571658} {\bibfield  {journal}
  {\bibinfo  {journal} {Journal of Mathematical Physics}\ }\textbf {\bibinfo
  {volume} {44}} (\bibinfo {year} {2003}),\ 10.1063/1.1571658}\BibitemShut
  {NoStop}%
\bibitem [{\citenamefont {Rigol}\ \emph {et~al.}(2007)\citenamefont {Rigol},
  \citenamefont {Dunjko}, \citenamefont {Yurovsky},\ and\ \citenamefont
  {Olshanii}}]{rigol_et_al_2007}%
  \BibitemOpen
  \bibfield  {author} {\bibinfo {author} {\bibfnamefont {M.}~\bibnamefont
  {Rigol}}, \bibinfo {author} {\bibfnamefont {V.}~\bibnamefont {Dunjko}},
  \bibinfo {author} {\bibfnamefont {V.}~\bibnamefont {Yurovsky}}, \ and\
  \bibinfo {author} {\bibfnamefont {M.}~\bibnamefont {Olshanii}},\ }\bibfield
  {title} {\enquote {\bibinfo {title} {{Relaxation in a Completely Integrable
  Many-Body Quantum System: An Ab Initio Study of the Dynamics of the Highly
  Excited States of 1D Lattice Hard-Core Bosons}},}\ }\href {\doibase
  10.1103/PhysRevLett.98.050405} {\bibfield  {journal} {\bibinfo  {journal}
  {Physical Review Letters}\ }\textbf {\bibinfo {volume} {98}},\ \bibinfo
  {pages} {050405} (\bibinfo {year} {2007})}\BibitemShut {NoStop}%
\bibitem [{\citenamefont {Vidmar}\ and\ \citenamefont
  {Rigol}(2016)}]{vidmar_rigol_2016}%
  \BibitemOpen
  \bibfield  {author} {\bibinfo {author} {\bibfnamefont {L.}~\bibnamefont
  {Vidmar}}\ and\ \bibinfo {author} {\bibfnamefont {M.}~\bibnamefont {Rigol}},\
  }\bibfield  {title} {\enquote {\bibinfo {title} {Generalized gibbs ensemble
  in integrable lattice models},}\ }\href {\doibase
  10.1088/1742-5468/2016/06/064007} {\bibfield  {journal} {\bibinfo  {journal}
  {Journal of Statistical Mechanics: Theory and Experiment}\ }\textbf {\bibinfo
  {volume} {2016}},\ \bibinfo {pages} {064007} (\bibinfo {year}
  {2016})}\BibitemShut {NoStop}%
\bibitem [{\citenamefont {Essler}\ and\ \citenamefont
  {Fagotti}(2016)}]{essler_fagotti_2016}%
  \BibitemOpen
  \bibfield  {author} {\bibinfo {author} {\bibfnamefont {F.~H.~L.}\
  \bibnamefont {Essler}}\ and\ \bibinfo {author} {\bibfnamefont
  {M.}~\bibnamefont {Fagotti}},\ }\bibfield  {title} {\enquote {\bibinfo
  {title} {Quench dynamics and relaxation in isolated integrable quantum spin
  chains},}\ }\href {\doibase 10.1088/1742-5468/2016/06/064002} {\bibfield
  {journal} {\bibinfo  {journal} {Journal of Statistical Mechanics: Theory and
  Experiment}\ }\textbf {\bibinfo {volume} {2016}},\ \bibinfo {pages} {064002}
  (\bibinfo {year} {2016})}\BibitemShut {NoStop}%
\bibitem [{\citenamefont {Peierls}(1929)}]{peierls_1929}%
  \BibitemOpen
  \bibfield  {author} {\bibinfo {author} {\bibfnamefont {R.}~\bibnamefont
  {Peierls}},\ }\bibfield  {title} {\enquote {\bibinfo {title} {Zur kinetischen
  theorie der w{\"a}rmeleitung in kristallen},}\ }\href
  {https://doi.org/10.1002/andp.19293950803} {\bibfield  {journal} {\bibinfo
  {journal} {Annalen der Physik}\ }\textbf {\bibinfo {volume} {395}},\ \bibinfo
  {pages} {1055--1101} (\bibinfo {year} {1929})}\BibitemShut {NoStop}%
\bibitem [{\citenamefont {Spohn}(2006{\natexlab{a}})}]{spohn_2006}%
  \BibitemOpen
  \bibfield  {author} {\bibinfo {author} {\bibfnamefont {H.}~\bibnamefont
  {Spohn}},\ }\bibfield  {title} {\enquote {\bibinfo {title} {{The Phonon
  Boltzmann Equation, Properties and Link to Weakly Anharmonic Lattice
  Dynamics}},}\ }\href {https://doi.org/10.1007/s10955-005-8088-5} {\bibfield
  {journal} {\bibinfo  {journal} {Journal of Statistical Physics}\ }\textbf
  {\bibinfo {volume} {124}},\ \bibinfo {pages} {1041–1104} (\bibinfo {year}
  {2006}{\natexlab{a}})}\BibitemShut {NoStop}%
\bibitem [{\citenamefont {Mendl}\ \emph {et~al.}(2016)\citenamefont {Mendl},
  \citenamefont {Lu},\ and\ \citenamefont {Lukkarinen}}]{mendl_et_al_2016}%
  \BibitemOpen
  \bibfield  {author} {\bibinfo {author} {\bibfnamefont {C.~B.}\ \bibnamefont
  {Mendl}}, \bibinfo {author} {\bibfnamefont {J.}~\bibnamefont {Lu}}, \ and\
  \bibinfo {author} {\bibfnamefont {J.}~\bibnamefont {Lukkarinen}},\ }\bibfield
   {title} {\enquote {\bibinfo {title} {{Thermalization of oscillator chains
  with onsite anharmonicity and comparison with kinetic theory}},}\ }\href
  {\doibase 10.1103/PhysRevE.94.062104} {\bibfield  {journal} {\bibinfo
  {journal} {Physical Review E}\ }\textbf {\bibinfo {volume} {94}},\ \bibinfo
  {pages} {062104} (\bibinfo {year} {2016})}\BibitemShut {NoStop}%
\bibitem [{\citenamefont {Lukkarinen}(2016)}]{lukkarinen_2016}%
  \BibitemOpen
  \bibfield  {author} {\bibinfo {author} {\bibfnamefont {J.}~\bibnamefont
  {Lukkarinen}},\ }\enquote {\bibinfo {title} {Kinetic theory of phonons in
  weakly anharmonic particle chains},}\ in\ \href {\doibase
  10.1007/978-3-319-29261-8_4} {\emph {\bibinfo {booktitle} {Thermal Transport
  in Low Dimensions: From Statistical Physics to Nanoscale Heat Transfer}}},\
  \bibinfo {editor} {edited by\ \bibinfo {editor} {\bibfnamefont
  {S.}~\bibnamefont {Lepri}}}\ (\bibinfo  {publisher} {Springer International
  Publishing},\ \bibinfo {address} {Cham},\ \bibinfo {year} {2016})\ pp.\
  \bibinfo {pages} {159--214}\BibitemShut {NoStop}%
\bibitem [{\citenamefont {Spohn}(2006{\natexlab{b}})}]{spohn_2006b}%
  \BibitemOpen
  \bibfield  {author} {\bibinfo {author} {\bibfnamefont {H.}~\bibnamefont
  {Spohn}},\ }\bibfield  {title} {\enquote {\bibinfo {title} {Collisional
  invariants for the phonon boltzmann equation},}\ }\href
  {https://doi.org/10.1007/s10955-006-9180-1} {\bibfield  {journal} {\bibinfo
  {journal} {Journal of Statistical Physics}\ }\textbf {\bibinfo {volume}
  {124}},\ \bibinfo {pages} {1131–1135} (\bibinfo {year}
  {2006}{\natexlab{b}})}\BibitemShut {NoStop}%
\bibitem [{\citenamefont {Mallayya}\ \emph {et~al.}(2019)\citenamefont
  {Mallayya}, \citenamefont {Rigol},\ and\ \citenamefont
  {De~Roeck}}]{mallayya_et_al_2019}%
  \BibitemOpen
  \bibfield  {author} {\bibinfo {author} {\bibfnamefont {K.}~\bibnamefont
  {Mallayya}}, \bibinfo {author} {\bibfnamefont {M.}~\bibnamefont {Rigol}}, \
  and\ \bibinfo {author} {\bibfnamefont {W.}~\bibnamefont {De~Roeck}},\
  }\bibfield  {title} {\enquote {\bibinfo {title} {{Prethermalization and
  Thermalization in Isolated Quantum Systems}},}\ }\href {\doibase
  10.1103/PhysRevX.9.021027} {\bibfield  {journal} {\bibinfo  {journal}
  {Physical Review X}\ }\textbf {\bibinfo {volume} {9}},\ \bibinfo {pages}
  {021027} (\bibinfo {year} {2019})}\BibitemShut {NoStop}%
\bibitem [{\citenamefont {Lenar\ifmmode \check{c}\else
  \v{c}\fi{}i\ifmmode~\check{c}\else \v{c}\fi{}}\ \emph
  {et~al.}(2018)\citenamefont {Lenar\ifmmode \check{c}\else
  \v{c}\fi{}i\ifmmode~\check{c}\else \v{c}\fi{}}, \citenamefont {Lange},\ and\
  \citenamefont {Rosch}}]{lenarcic_et_al_2018}%
  \BibitemOpen
  \bibfield  {author} {\bibinfo {author} {\bibfnamefont {Z.}~\bibnamefont
  {Lenar\ifmmode \check{c}\else \v{c}\fi{}i\ifmmode~\check{c}\else
  \v{c}\fi{}}}, \bibinfo {author} {\bibfnamefont {F.}~\bibnamefont {Lange}}, \
  and\ \bibinfo {author} {\bibfnamefont {A.}~\bibnamefont {Rosch}},\ }\bibfield
   {title} {\enquote {\bibinfo {title} {Perturbative approach to weakly driven
  many-particle systems in the presence of approximate conservation laws},}\
  }\href {\doibase 10.1103/PhysRevB.97.024302} {\bibfield  {journal} {\bibinfo
  {journal} {Physical Review B}\ }\textbf {\bibinfo {volume} {97}},\ \bibinfo
  {pages} {024302} (\bibinfo {year} {2018})}\BibitemShut {NoStop}%
\bibitem [{\citenamefont {Lange}\ \emph {et~al.}(2018)\citenamefont {Lange},
  \citenamefont {Lenar\ifmmode \check{c}\else
  \v{c}\fi{}i\ifmmode~\check{c}\else \v{c}\fi{}},\ and\ \citenamefont
  {Rosch}}]{lange_et_al_2018}%
  \BibitemOpen
  \bibfield  {author} {\bibinfo {author} {\bibfnamefont {F.}~\bibnamefont
  {Lange}}, \bibinfo {author} {\bibfnamefont {Z.}~\bibnamefont {Lenar\ifmmode
  \check{c}\else \v{c}\fi{}i\ifmmode~\check{c}\else \v{c}\fi{}}}, \ and\
  \bibinfo {author} {\bibfnamefont {A.}~\bibnamefont {Rosch}},\ }\bibfield
  {title} {\enquote {\bibinfo {title} {{Time-dependent generalized Gibbs
  ensembles in open quantum systems}},}\ }\href {\doibase
  10.1103/PhysRevB.97.165138} {\bibfield  {journal} {\bibinfo  {journal}
  {Physical Review B}\ }\textbf {\bibinfo {volume} {97}},\ \bibinfo {pages}
  {165138} (\bibinfo {year} {2018})}\BibitemShut {NoStop}%
\bibitem [{\citenamefont {Reimann}\ and\ \citenamefont
  {Dabelow}(2019)}]{reimann_dabelow_2019}%
  \BibitemOpen
  \bibfield  {author} {\bibinfo {author} {\bibfnamefont {P.}~\bibnamefont
  {Reimann}}\ and\ \bibinfo {author} {\bibfnamefont {L.}~\bibnamefont
  {Dabelow}},\ }\bibfield  {title} {\enquote {\bibinfo {title} {Typicality of
  prethermalization},}\ }\href {\doibase 10.1103/PhysRevLett.122.080603}
  {\bibfield  {journal} {\bibinfo  {journal} {Physical Review Letters}\
  }\textbf {\bibinfo {volume} {122}},\ \bibinfo {pages} {080603} (\bibinfo
  {year} {2019})}\BibitemShut {NoStop}%
\bibitem [{Note1()}]{Note1}%
  \BibitemOpen
  \bibinfo {note} {This value may also be recovered from a simple power
  counting argument: Let $\protect \mathrm {ad}_H =\protect \{H, \cdot \protect
  \}$ ($= -i[ H, \cdot ]$ for a quantum system with $\hbar =1$) and expand
  ${\protect \mathrm e}^{-t\protect \mathrm {ad}_H} N_0$ at order $p$; this
  yields $(-t)^p\protect \mathrm {ad}_H^p (N_0) / p!$ which is a polynomial of
  order $n$ in $a^\pm $, with $n$ as given.}\BibitemShut {Stop}%
\end{thebibliography}%

\onecolumngrid

\pagebreak

\begin{center}
\LARGE{
Supplemental material
}
\end{center}

\section{Solving eq.~(\ref{eq: resonant constraints})}

Here we derive bounds on the quantity $\delta_c (n)$, with $n\ge 2$ even, such that eq.~(\ref{eq: resonant constraints}) only admits solutions for $\delta \ge \delta_c (n)$. 
The nearest neighbor dispersion relation is given by eq.~(\ref{eq: dispersion relation}), i.e.\@ 
$$
	\omega(k) := \omega_0 \sqrt{1-2\delta \cos(2\pi k)}\,
$$
with $\omega_0>0$, $0<\delta\le \frac{1}{2}$.
Eq.~(\ref{eq: resonant constraints}) admits a solution if the number non-preserving collisional manifold is not empty, 
i.e. if there is $k \in (\mathrm{BZ})^n$ and $\sigma\in \{ \pm 1\}^n$
for which
\begin{equation}\label{eq:collconstr}
	\Omega(k,\sigma) := \sum_{\ell=1}^n \sigma_\ell \omega(k_\ell) =0 , \qquad
   	\sum_{\ell=1}^n k_\ell =0 \quad (\text{modulo }1),
\end{equation}
with $\sigma$ such that $\Delta N_0 := \sum_{\ell=1}^n \sigma_\ell\ne 0$ (we consider only processes that do not preserve the number of phonons). 

To clearly make the connection with eq.~(\ref{eq: resonant constraints}), we notice that
by sign-change symmetry, we may focus on the case with $\Delta N_0 >0$, permute the labels so that all positive signs come before the negative ones, 
and remark that eq.~\eqref{eq:collconstr} admits a solution for $\Delta N_0 >0$ if and only if it admits a solution for $\Delta N_0 = 2$.
This last point follows from the fact that $\Omega (k, \sigma) \ge \Omega (k, \sigma')$ for any $k \in (\mathrm{BZ})^n$,
if $\sigma_\ell \ge \sigma'_\ell$ for all $1 \le \ell \le n$, 
and from the fact that eq.~(\ref{eq:collconstr}) admits no solution 
if and only if $\Omega(k,\sigma)$ does not change sign, i.e.\@ remains strictly positive/negative, 
on the set $\{k \in (\mathrm{BZ})^n : \sum_{\ell = 1}^n k_\ell = 0\} \simeq (\mathrm{BZ})^{n-1}$
(and the sign is the sign of $\Delta N_0$ since $\Omega (0,\sigma)$ is proportional to $\Delta N_0$).  
Below, for simplicity, we set $\omega_0=1$ since its value will not affect the value of $\delta_c$.

Denote the minimum of $\omega (k)$ by $m_-$ and maximum by $m_+ $.  The minimum is reached at $k=0$ and the maximum at $k=\frac{1}{2}$ and thus
$$
 	m_- = \sqrt{1-2\delta},\qquad m_+ = \sqrt{1+2\delta}.
$$
Denote $n_{\pm} = \#\{\ell: \sigma_\ell=\pm 1\}$ for which obviously $n=n_+ + n_-$ and $\Delta N_0 = n_+-n_-\ge 2$.
We then have
$$
 	\Omega(k) \ge n_+ m_- - n_- m_+.
$$
Since here $n_\pm = (n\pm \Delta N_0)/2$, it follows that
\begin{equation}\label{eq: lower bound on Omega}
 	2\Omega(k) \ge \Delta N_0 (m_+ +m_-)-n (m_+- m_-) . 
\end{equation}

Consider then the following function of $\delta\in {]}0,\frac{1}{2}]$:
$$
 	G(\delta) := 2\frac{m_+ +m_-}{m_+ -m_-}
 	= 
 	2 \frac{\sqrt{1+2\delta}+\sqrt{1-2\delta}}{\sqrt{1+2\delta}-\sqrt{1-2\delta}}
 	= 
	\frac{1+\sqrt{1-4\delta^2}}{\delta}.
$$
Clearly,
\begin{equation}\label{eq:lowewGbound}
 	\Delta N_0 (m_+ +m_-) - n (m_+- m_-)
 	= 
	(\Delta N_0 -2)(m_+ +m_-) + (G(\delta)-n) (m_+- m_-)
\end{equation}
Here $G$ is strictly decreasing from $+\infty$ to $2$, and thus there are unique values $\delta_-(N)$ obtained as a solutions of 
$$
 	G(\delta_-(n))=n .
$$
A computation yields 
\begin{equation}\label{eq: delta - n}
	\delta_- (n) = \frac{2n}{n^2 + 4}
\end{equation}
and in particular, $n \delta_-(n) \to 2$ as $n\to \infty$.
Since $G(\delta)>n$ if and only if $\delta <\delta_-(n)$, it follows from eq.~(\ref{eq: lower bound on Omega}-\ref{eq:lowewGbound}) 
that $\Omega(k)>0$ for all $k$ if $\delta <\delta_-(n)$, and thus $\delta_c(n)\ge \delta_-(n)$ for all $n\ge 2$.
In particular, $\delta_c(2)=\frac{1}{2}$.

In the above estimates, we have not used the translation invariance constraint in eq.~\eqref{eq:collconstr}, $\sum_\ell k_\ell =0$, at all.  
In particular, it plays an important role in the case $n=4$, $\Delta N_0 = 2$:
As shown in \cite[Sec.\ 2.2]{lukkarinen_2016}, there is a constant $C_\delta>0$ such that $|\Omega|\ge C_\delta >0$ for $\delta < \frac12$,
whenever $\sum_\ell k_\ell=0$ modulo one.
Here one may use for instance $C_\delta=\frac{m_-}{2}\text{arcosh}\frac{1}{2\delta}$  
which goes to zero as $\delta \to \frac{1}{2}$ but otherwise is strictly bounded away from zero (note that $\omega$ is symmetric under $k\mapsto -k$).  
Therefore, we also have $\delta_c(4)=\frac{1}{2}$. 
Note that this bound is an improvement of the earlier bound which had $\delta_-(4)=0.4$.

Consider next $n\ge 6$ such that $n/2$ is odd.
As explained earlier, to show that there is a solution to eq.~\eqref{eq:collconstr}, 
it is enough to find a value of $k$ satisfying the translation invariance constraint and such that $\Omega (k, \sigma) \le 0$ for $\Delta N_0 = 2$.  
Choose $k=k^0$, where $k^0_\ell=0$ when $\sigma_\ell=+1$, and $k^0_\ell=\frac{1}{2}$ when $\sigma_\ell=-1$.  
As in this case $n_- = (n-\Delta N_0)/2$ is even, 
the sum $ \sum_{\ell=1}^n k^0_\ell$ yields an integer and thus $k^0$ satisfies the translation invariance constraint.  
On the other hand, $\Omega(k^0)=n_+ m_- - n_- m_+ = \frac{1}{2}(G(\delta)-n) (m_+- m_-)$ and thus if $\delta>\delta_-(n)$, we have $\Omega(k^0)<0$.  
Therefore, in this case there is a solution to (\ref{eq:collconstr}).
We can conclude that, if $n/2$ is odd, then  $\delta_c(n)=\delta_-(n)$.

Finally, let us consider $n\ge 8$ such that $n/2$ is even. 
Define $k^0$ as above, and note that then 
$\sum_\ell k^0_\ell = \frac{n}{4}-\frac{1}{2}$.  
Set 
$\tilde{k}^0_\ell:= k^0_\ell-\frac{1}{2n}$ for which $\sum_\ell \tilde{k}^0_\ell=0$ modulo $1$, and thus the translation invariance constraint is satisfied.  
On the other hand,
\begin{align*}
 	\Omega(\tilde{k}^0) 	
 	&= 
 	n_+ \omega\left(\frac{1}{2n}\right) - n_- \omega\left(\frac{1}{2}-\frac{1}{2n}\right)\\
 	&=
	n_+ \sqrt{1-2\delta \cos(\pi /n)}- n_- \sqrt{1+2\delta \cos(\pi /n)}
  	=
	n_+ m_-(\delta') - n_- m_+(\delta')
	=
 	\frac{1}{2}(G(\delta')-n) \big(m_+(\delta') - m_-(\delta')\big)  .
\end{align*}
with $\delta'= \delta \cos(\pi /n)$ and $m_\pm (\delta') = \sqrt{1 \pm 2 \delta'}$.
Now, if $\delta>\delta_-(n)/\cos(\pi /n)$, we have $G(\delta')<n$,
and thus $\Omega(\tilde{k}^0)<0$ and  there exists a solution to (\ref{eq:collconstr}).
We can conclude that, if $n\ge 8$ and $n/2$ is even, then $\delta_-(n)\le \delta_c(n) \le \delta_-(n)/\cos(\pi /n)$.
In particular, also in this case $n\delta_c(n)\to 2$ as $n\to \infty$.  

For $n=8$, the above bounds yields $0.235<\delta_c(8)<0.255$, 
and further numerical checks of the values of $\Omega$ on the values satisfying the translation invariance constraint show that $\delta_c(8)\approx 0.25$.

\section{Proof of Claim~\ref{cl: new variables} and Claim~\ref{cl: dressed quantity}}

We provide here a rigorous proof of Claim~\ref{cl: new variables} and Claim~\ref{cl: dressed quantity}. 
Let us first deal with Claim~\ref{cl: new variables}.
Let $r$ and $\delta$ be fixed such that $p > 1$.
Let $L$ be the length of the chain, and let us assume that the Hamiltonian in eq.~(\ref{eq: hamiltonian}) is defined with periodic boundary conditions.
Eq.~(\ref{eq: phonon field}-\ref{eq: V}) still make sense, provided that we define 
$\int_{\mathrm{BZ}} \dd k \hat\varphi (k) = \frac{1}{L} \sum_{k=0}^{L-1} \hat\varphi (k/L)$ 
with $\hat\varphi (k) = \sum_{x=1}^L \varphi (x) \ed^{- 2i \pi k x}$, 
and $\delta (k/L) = L$ for $k=0$ and $\delta (k/L) = 0$ otherwise. 
Let finally the Poisson bracket for two functions $f,g$ on the phase space $\R^{2L}$ be defined as
\begin{equation}\label{eq: Poisson bracket}
	\mathrm{ad}_f(g) = \{ f,g \} = \nabla_q f \cdot \nabla_p g - \nabla_p f \cdot \nabla_q g. 
\end{equation}

Let us first perform formal computations that we will justify afterwards. 
Given a function $-G = \sum_{n=1}^{p-1} \lambda^n G_n$ on the phase space, we can expand the operator $\ed^{- \mathrm{ad}_G} = \sum_{n\ge 0} \lambda^n S_n$
with $S_0 = \mathrm{Id}$ and 
\begin{equation}\label{eq: first expansion}
	S_n
	= 
	\sum_{m=1}^n \frac{1}{m!} \sum_{\substack{1 \le k_1, \dots , k_m < p, \\ k_1 + \dots + k_m = n}} \mathrm{ad}_{G_{k_1}} \dots \mathrm{ad}_{G_{k_m}}, 
	\qquad n\ge 1.
\end{equation}
For $n\ge 1$, we further decompose $S_n$ as $S_n = \mathrm{ad}_{G_n} + T_{n-1}$ and we notice that $T_n$ only involves coefficients $G_k$ with $k \le n$.
Hence 
\begin{equation}\label{eq: change of variables expansion}
	\ed^{- \mathrm{ad}_G} H
	= 
	H_0 + 
	\sum_{n\ge 1} \lambda^n \big( S_n H_0 + S_{n-1}V \big)
	= 
	H_0 + 
	\sum_{n\ge 1} \lambda^n \big(  \{ G_n , H_0 \} +  T_{n-1} H_0 + S_{n-1}V \big).
\end{equation}

For $m \ge 2$, let us consider functions as in eq.~(\ref{eq: order n observable}), 
i.e.\@ translation invariant homogeneous polynomials (TIHP) of order $n$: 
\begin{equation}\label{eq: order n polynomial}
	\varphi_m
	= 
	\int_{(\mathrm{BZ})^m} \dd k_1 \dots \dd k_m \delta(k_1 + \dots + k_m) 
	\sum_{\sigma_j = \pm} \hat\varphi_m(k_1,\dots,k_m,\sigma_1, \dots ,\sigma_m)a_{1}^{\sigma_1} \dots a_m^{\sigma_m}
\end{equation}
where $\hat\varphi_m$ is analytic on $(\mathrm{BZ})^m$. 
Translation invariant polynomials (TIP) of order $m\ge 2$ are functions of the form $f_m = \sum_{k=2}^m \varphi_k$ where $\varphi_k$ are TIHPs of order $k$. 
If $\varphi$ is a TIHP, it can be decomposed as $\varphi = \varphi_\parallel + \varphi_\perp$, 
where $\varphi_\parallel$ collects the terms such that $\sum_{j=1}^m\sigma_j = 0$ in \eqref{eq: order n polynomial}. 
TIPs can be decomposed accordingly.
The crucial property implied by this decomposition is that $\mathrm{ad}_{N_0} (\varphi_\parallel) = 0$. 
Assuming that our computations involve only TIPs, and we will show below that this assumption is legitimate, 
we can find the coefficients $G_n$ so that $\{ \ed^{-\mathrm{ad}_G H}, N_0\} = \mathcal O (\lambda^p)$.
For this, we require that $G_n$ solve the set of recursive equations
\begin{equation}\label{eq: recursive equations Gn}
	\{ H_0, G_n \} = (T_{n-1} H_0 + S_{n-1}V)_\perp, \qquad 1 \le n \le p-1.
\end{equation}
Indeed, inserting \eqref{eq: recursive equations Gn} in \eqref{eq: change of variables expansion} yields
\begin{equation}\label{eq: H new variables}
	\ed^{- \mathrm{ad}_G} H
	=
	H_0 + 
	\sum_{n=1}^{p-1} \lambda^n(T_{n-1} H_0 + S_{n-1}V)_\parallel + \lambda^{p} \sum_{n \ge p} \lambda^{n-p} \big(T_{n-1} H_0 + S_{n-1}V \big) 
\end{equation}
and thus 
\begin{equation}\label{eq: evaluation of the current new variables}
	\{ N_0 , \ed^{- \mathrm{ad}_G} H \}
	=
	\mathrm{ad}_{N_0} (\ed^{- \mathrm{ad}_G} H)
	=
	\lambda^{p} \sum_{n \ge p} \lambda^{n-p} \big\{ N_0, ( T_{n-1} H_0 + S_{n-1}V ) \big\} .
\end{equation}

To show that the above scheme make sense, and establish Claim 1, we need to prove that the equations \eqref{eq: recursive equations Gn} can be solved, 
and that the expansion \eqref{eq: H new variables} converges for $|\lambda|$ small enough. 
Let us start with eq.~\eqref{eq: recursive equations Gn}.
From the definitions \eqref{eq: Poisson bracket} and  (\ref{eq: phonon field}), we derive the canonical commutation rule 
\begin{equation}\label{eq: CCR}
	\{ a^\sigma (k) , a^{\sigma'} (k') \} = i \sigma \delta_{\sigma + \sigma'} \delta (k + k'). 
\end{equation}
Together with the rule $\{ f, gh \} = \{ f ,g \} h + g \{ f,h\}$,
we can readily evaluate the Poisson bracket between TIHPs. 
In particular we derive that if $\varphi_{n_1}$ is a TIHP of order $n_1$ and if $\varphi_{n_2}$ is a TIHP of order $n_2$, 
then $\{ \varphi_{n_1},\varphi_{n_2}\}$ is a TIHP of order $n_1+n_2-2$.
Moreover, if $\varphi_n$ is a TIHP of order $n$ with kernel $\hat\varphi_n (k_1, \dots , k_n, \sigma_1, \dots , \sigma_n)$, 
then $\{ H_0 , \varphi_n\}$ is again a TIHP of order $n$ with kernel
$$
	-i \left(\sum_{j=1}^n \sigma_j \omega_j \right) \hat\varphi_n(k_1,\dots,k_n,\sigma_1, \dots ,\sigma_n) .
$$
Hence, if $\varphi_n$ is a TIHP of order $n$, and if $\delta < \delta_c (n)$, ensuring that eq.~(\ref{eq: resonant constraints}) has no solution, 
then the equation 
$$
	\{ H_0 , u \} = (\varphi_n)_\perp
$$
admits a solution $u$ given by 
\begin{equation}\label{eq: solution poisson equation}
	u = 
	i 
	\int_{(\mathrm{BZ})^n} \dd k_1 \dots \dd k_n \delta(k_1 + \dots + k_n) 
	\sum_{\sigma_j = \pm} \frac{\delta (\sum_{j=1}^n \sigma_j \ne 0)}{\sum_{j=1}^n \sigma_j \omega_j} 
	\hat\varphi_n(k_1,\dots,k_n,\sigma_1, \dots ,\sigma_n)a_{1}^{\sigma_1} \dots a_n^{\sigma_n}
\end{equation}
with the convention $0/0 = 0$. 

If we first do not pay attention to the regularity of the kernels involved, 
i.e.\@ if we ignore possible singularities stemming from the fact that $\sum_{j=1}^n \sigma_j \omega_j$ may vanish in eq.~\eqref{eq: solution poisson equation}, 
we find that $G_n$ solving eq.~\eqref{eq: recursive equations Gn} are TIPs of order $nr - 2(n-1)$.
To show next that singularities do not occur and that the kernels are analytic, we use that $\delta$ and $p$ satisfy eq.~(\ref{eq: p value}). 
This guarantees in particular that 
$$
	\delta < \delta_c ((p-1)(r-2)+2) \le \delta_c (nr - 2(n-1)) \quad \text{for all} \quad 1 \le n \le p-1, 
$$
and therefore analyticity. 
Finally, even though this is not needed for the proof, 
we notice also that we do not expect to be able to find a regular function $G_p$ solving eq.~\eqref{eq: recursive equations Gn}, 
since $\delta > \delta_c (p(r-2)+2)$ by eq.~(\ref{eq: p value}), 
i.e.\@ we expect to have reached the optimal order $p$. 

We next deal with the convergence of the expansion in \eqref{eq: H new variables}. 
Let us consider the Hamiltonian $H$ on the extended space $\R^{2L + 1}$, so as to explicitly include the dependence of $H$ on $\lambda$, 
and let us consider the Cauchy problem 
\begin{equation}\label{eq: pde}
	\partial_\tau \tilde H(\tau,x) = - \{ G , \tilde H \} (\tau,x) \quad (x \in \R^{2L + 1}), \qquad \tilde H(0,x) = H(x).
\end{equation}
We observe that $\tilde H(1,\cdot) = \ed^{-\mathrm{ad}_G}H$, if both terms make sense. 
By the Cauchy--Kowalevski theorem, eq.~\eqref{eq: pde} admits a real analytic solution in the neighborhood of the origin in $\R \times \R^{2L +1}$. 
Moreover, since $G = \mathcal O (\lambda)$, we may assume that the solution is well defined up to $\tau=1$ by shrinking the neighborhood in $\lambda$. 
This ensures the convergence of eq.~\eqref{eq: H new variables}.

Let us finally move to Claim~\ref{cl: dressed quantity}. 
We notice that $\ed^{\mathrm{ad}_G} \{ \ed^{-\mathrm{ad}_G} H, N_0 \} = \{ H , \ed^{\mathrm{ad}_G}N_0 \}$,
where both sides of the equality are well defined and analytic in $\lambda$ in a neighborhood of the origin, by a similar argument as before.
Moreover, $\{ \ed^{-\mathrm{ad}_G} H, N_0 \} = \mathcal O (\lambda^p)$ by our construction, 
hence also $\ed^{\mathrm{ad}_G} \{ \ed^{-\mathrm{ad}_G} H, N_0 \} = \mathcal O (\lambda^p)$. 
Writing $\ed^{\mathrm{ad}_G}N_0 = \sum_{n\ge 0} \lambda^n N_n$ and defining $N = \sum_{n=0}^{p-1} \lambda^n N_n$, 
we conclude that $\{ H , N \} = \lambda^p \{ N_{p-1} , V\}$. 
The quantity $N$ defines an extensive quantity in the thermodynamic limit $L \to \infty$, and the last relation remains true in this limit. 
This yields thus  Claim~\ref{cl: dressed quantity}.

\section{Derivation of the dissipation rate: eq.~(\ref{eq: rate p=1}) and eq.~(\ref{eq: rate p>1})}

We derive the expressions for the dissipation rate $\gamma$ in eq.~(\ref{eq: rate p=1}), valid for $p=1$, and eq.~(\ref{eq: rate p>1}), valid for $p>1$. 

\bigskip
\noindent
\textbf{Eq.~(\ref{eq: rate p=1}): $p=1$ ---}
Let a Wigner function $W$ be given, and let us assume that the systems is in a state of the form 
$$
	\rho = \frac{1}{Z} \rho_0 (1 + \lambda f + \mathcal O (\lambda^2))
	\qquad \text{with} \qquad 
	\rho_0 =  \frac{1}{Z_0}\ed^{- \int_{\mathrm{BZ}} \dd k  \frac{n(k)}{W(k)}}
$$
where $\lambda f$ represents a first order correction. 
This assumption is the analog of the assumption (\ref{eq: main assumption}) that will be used for the case $p>1$. 
If $\varphi$ is any observable of the type $\varphi = \int_{\mathrm{BZ}} \dd k \hat\varphi (k) n(k)$, 
its flux $J_\varphi = \{ H,\varphi \} = \lambda \{ V,\varphi\} = \lambda \mathcal J_\varphi$
vanishes on average in the state $\rho_0$: $\langle J_\varphi \rangle_{\rho_0} = 0$. 
Therefore, the occupations of all phonon modes $n(k)$ must evolve on time scales of order $\lambda^2$, 
and the whole state $\rho$ evolves thus only on these time scales. 
Expressing this mathematically determines the first order correction $f$: 
\begin{equation}\label{eq: condition on f}
	\mathrm{ad}_H^\dagger (1 + \lambda f) = \mathcal O (\lambda^2)
\end{equation}
where $\mathrm{ad}^\dagger_H$ is the adjoint of $\mathrm{ad}_H$ with respect to the measure $\rho_0$.

Let us compute the adjoint $\mathrm{ad}^\dagger_H$. 
It is defined as the operator such that $\int \rho_0 	(\mathrm{ad}_H u ) v = \int\rho_0 u (\mathrm{ad}^\dagger_H v)$ for any functions $u,v$.
We compute
$$
	\int (\mathrm{ad}_H u) v \rho_0 
	= 
	- \int u \mathrm{ad}_H (v \rho_0)
	= 
	\int u (-\mathrm{ad}_H v) \rho_0 + u v (- \mathrm{ad}_H \rho_0)
$$
and
$$
	- \mathrm{ad}_H \rho_0
	= 
	\rho_0 \left\{ H,  \int \frac{n(k)}{W(k)} \, \dd k \right\}
	= 
	\lambda \rho_0 \left\{ V , \int \frac{n(k)}{W(k)} \, \dd k \right\}
	=
	\lambda \rho_0 \mathcal J_{1/W}.
$$
Therefore
\begin{equation}\label{eq: adjoint H}
	\mathrm{ad}^\dagger_H  v = - \mathrm{ad}_H v + \lambda \mathcal J_{1/W} .
\end{equation}

Combining eq.~(\ref{eq: condition on f}) and eq.~(\ref{eq: adjoint H}), we find that $f$ satisfies $\{ H_0 , f \} = \mathcal J_{1/W}$.
However, due to resonances, i.e.\@ due to the fact that $0$ is in the spectrum of $\mathrm{ad}_{H_0}$, we insert a regularization to solve this equation: 
Given $\tau < + \infty$, we consider instead the equation $(\mathrm{ad}_{H_0} + \frac{1}{\tau}) f = g$ and consider the limit $\tau \to \infty$. 
This can be solved as 
\begin{equation}\label{expression for f}
	f  
	= 
	\Big(\mathrm{ad}_{H_0} + \frac{1}{\tau}\Big)^{-1} \mathcal J_{1/W}
	=
	\int_0^\infty \dd t\, \ed^{- t /\tau} \mathcal J_{1/W}(t), \quad \tau \to \infty,
\end{equation}
where $g(t)$ is the evolution of $g$ for the free dynamics generated by $\mathrm{ad}_{H_0}$.

We can now derive eq.~(\ref{eq: rate p=1}).
Let $J = \lambda \mathcal J = \lambda \{ V,N_0 \}$. We compute
\begin{equation}\label{eq: gamma W appendix}
	\gamma (W)
	=
	\frac{\langle J\rangle_\rho}{\delta (0)}
	= 
	\frac{\lambda}{\delta(0)} \langle \mathcal J \rangle_\rho
	= 
	\frac{\lambda^2}{\delta(0)} \langle \mathcal J f \rangle_{\rho_0} + \mathcal O (\lambda^3)
	= 
	\frac{\lambda^2}{\delta(0)} \lim_{\tau\to \infty} \int_0^\infty \dd t \, \ed^{- t/\tau} \langle \mathcal J (0) \mathcal J_{1/W} (t) \rangle_{\rho_0} 
	+ \mathcal O (\lambda^3) . 
\end{equation}

\bigskip
\noindent
\textbf{Eq.~(\ref{eq: rate p>1}): $p>1$ ---}
We proceed in a very similar way. 
As explained in the main text, we find it convenient to move to the rotated frame where $N_0$ is a pseudo-conserved quantity for a dressed Hamiltonian $\tilde H$. 
According to eq.~(\ref{eq: main assumption}), we assume that the system is in a state $\rho$ of the form 
$$
	\rho = \frac{1}{Z}\rho_0 (1 + \lambda^p f + \mathcal O (\lambda^{p+1})) 
	\quad \text{with} \quad
	\rho_0 = \frac{1}{Z_0} \ed^{- \beta (\tilde H - \mu N_0)}  
$$
where $\lambda^p f$ represents the correction at order $p$. 
As derived in the main text, the flux of $N_0/L$, i.e. $J = \{ \tilde H, N_0 \} = \lambda^p \mathcal J$ vanishes in the state $\rho_0$: $\langle J \rangle_{\rho_0} = 0$.   
Hence, since $N_0$ is the only quantity that brings the system out of equilibrium,
the evolution of the whole state $\rho$ must itself occur on time scales of order $\lambda^{2p}$. 
This yields in particular a relation analogous to eq.~(\ref{eq: condition on f}):
\begin{equation}\label{eq: condition on f for p>1}
	\mathrm{ad}_{\tilde H}^\dagger (1 + \lambda^p f) = \mathcal O (\lambda^{p+1}) 
\end{equation}
where $\mathrm{ad}_{\tilde H}^\dagger$ is the adjoint with respect to $\rho_0$. 
Again, we compute that this operator acts on a function $v$ as: 
\begin{equation}\label{eq: adjoint H tilde}
	\mathrm{ad}_{\tilde H}^\dagger v 
	= 
	- \mathrm{ad}_{\tilde H} v - \beta\mu \{ \tilde H, N_0 \} 
	=
	- \mathrm{ad}_{\tilde H} v - \lambda^p\beta\mu\mathcal J .
\end{equation}
Thus, combining eq.~(\ref{eq: condition on f for p>1}) and eq.~(\ref{eq: adjoint H tilde}), 
we derive that $f$ must satisfy $\mathrm{ad}_{\tilde H} f = - \beta \mu \mathcal J$ in lowest order in $\lambda$, i.e. $\mathrm{ad}_{H_0} f = - \beta \mu \mathcal J$. 
Again, this equation needs to be regularized, and we get 
\begin{equation}\label{expression for f for p > 1}
	f  
	= 
	-\beta \mu \Big(\mathrm{ad}_{H_0} + \frac{1}{\tau}\Big)^{-1} \mathcal J
	=
	-\beta \mu \int_0^\infty \dd t\, \ed^{- t /\tau} \mathcal J(t), \quad \tau \to \infty,
\end{equation}
where, again, $\mathcal J(t)$ is the evolution of $\mathcal J$ under the free dynamics generated by $H_0$.  
We come to the conclusion that 
\begin{equation}\label{eq: gamma beta mu appendix}
	\gamma (\beta,\mu)
	=
	\frac{\langle J\rangle_\rho}{\delta (0)}
	= 
	\frac{\lambda^p}{\delta(0)} \langle \mathcal J \rangle_\rho
	= 
	\frac{\lambda^{2p}}{\delta(0)} \langle \mathcal J f \rangle_{\rho_0} + \mathcal O (\lambda^{2p+1})
	= 
	-\frac{\beta \mu \lambda^{2p}}{\delta(0)} \lim_{\tau\to \infty} \int_0^\infty \dd t \, \ed^{- t/\tau} \langle\mathcal J (0) \mathcal J (t)\rangle_{\rho_0} 
	+ \mathcal O (\lambda^{2p+1}) . 
\end{equation}

\section{Explicit evaluation of the dissipation rate in specific cases}

We compute explicitly the dissipation rate $\gamma$ in the leading order in $\lambda$ for the three cases where we want to compare our predictions with numerical data. 
Our starting point is always the expression \eqref{eq: gamma beta mu appendix} 
(even for $p=1$, since if we take $W = (\beta(\omega(k) - \mu)^{-1}$, the expressions \eqref{eq: gamma W appendix} and \eqref{eq: gamma beta mu appendix} coincide). 
Eq.~\eqref{eq: gamma beta mu appendix} still contains some hidden dependence in $\lambda$ through $\mathcal J$ and $\rho_0$. 
To obtain the leading order, we replace $\langle \cdot \rangle_{\rho_0}$ 
by the average $\langle \cdot \rangle$ over a Gaussian measure with density $\ed^{- \beta (H_0 - \mu N_0)}/Z_0$. 
Omitting $\mathcal O (\lambda^{2p + 1})$ terms in our formulas for simplicity, and writing $\varepsilon = \tau^{-1}$, we get 
\begin{equation}\label{eq: rate gamma appendix}
	\gamma 
	= 
	-\frac{\beta \mu \lambda^{2p}}{\delta (0)} \lim_{\varepsilon \to 0} \langle \mathcal J (\mathrm{ad}_{H_0} + \varepsilon)^{-1} \mathcal J \rangle .
\end{equation}

\bigskip
\underline{$r=6$ and $\delta > 0.3$:} 
In this case $p=1$.
Our aim is to show that 
\begin{equation}\label{eq: r=6 delta > 0.3}
	\gamma = \gamma_0 \lambda^2 \beta^{-5} \quad \text{for} \quad \mu = -1
\end{equation}
where $\gamma_0$ is a number that depends only on the value of $\delta$ and that can be evaluated explicitly. 

Since $\mathcal J = \{ V,N_0 \}$, eq.~\eqref{eq: rate gamma appendix} becomes 
$$
	\gamma \; = \; - \frac{\beta \mu \lambda^2}{\delta(0)} \lim_{\varepsilon \to 0} \langle \{ V, N_0 \} (\mathrm{ad}_{H_0} + \varepsilon)^{-1} \{ V, N_0 \} \rangle .
$$
A computation yields
$$
	\{ V , N_0 \}
	\; = \; 
	\frac{i}{48} \int \frac{\dd k_1 \dots \dd k_6}{(\omega_1 \dots \omega_6)^{1/2}} \delta(k_1 + \dots + k_6)
	\sum_{\sigma_i} (\sigma_1 + \dots + \sigma_6) a_1^{\sigma_1} \dots a_6^{\sigma_6}
$$
Next, to obtain $(\mathrm{ad}_{H_0} + \varepsilon)^{-1} \{ V, N_0 \}$, we compute that 
\begin{equation}\label{eq: inversion 1}
	(\mathrm{ad}_{H_0} + \varepsilon)^{-1} a_1^{\sigma_1} \dots a_6^{\sigma_6} 
	= 
	\frac{1}{- i (\sigma_1 \omega_1 + \dots + \sigma_6 \omega_6) + \varepsilon} a_1^{\sigma_1} \dots a_6^{\sigma_6} 
\end{equation}
We anticipate that, because of cancellations, only the real part of the fraction on the right hand side brings a non-zero contribution, and we compute
\begin{equation}\label{eq: inversion 2}
	\lim_{\varepsilon \to 0} \Re \frac{1}{- i (\sigma_1 \omega_1 + \dots + \sigma_6 \omega_6) + \varepsilon} = \pi \delta (\sigma_1 \omega_1 + \dots + \sigma_6 \omega_6) . 
\end{equation}
Hence,
\begin{align*}
	\gamma \: =& \; -\frac{\beta \mu \lambda^2}{\delta (0)}
	\frac{-\pi}{(48)^2} \int \frac{\dd k_1 \dots \dd k_{12}}{(\omega_1 \dots \omega_{12})^{1/2}} \delta(k_1 + \dots + k_6)\delta(k_7 + \dots + k_{12})\\
	&\times\sum_{\sigma_i} \delta(\sigma_7 \omega_7 + \dots + \sigma_{12} \omega_{12})
	(\sigma_1 + \dots + \sigma_6)(\sigma_7 + \dots + \sigma_{12}) \langle a_1^{\sigma_1} \dots a_{12}^{\sigma_{12}} \rangle \, .
\end{align*}

We now must expand $\langle a_1^{\sigma_1} \dots a_{12}^{\sigma_{12}} \rangle$ by performing Gaussian pairings with the rule 
$$
	\langle a^{\sigma_i}(k_i) a^{\sigma_j} (k_j) \rangle = \frac{\delta(\sigma_i + \sigma_j)}{\beta (\omega_j - \mu)}.
$$
We see that we must pair variables with indices $1, \dots , 6$ to variables $7, \dots , 12$, 
as otherwise one would be left with terms involving only four phonons, and these vanish 
since $(\sigma_1 + \dots + \sigma_4) \delta (\sigma_1 \omega_1 + \dots + \sigma_4 \omega_4) = 0$.
There are $6!$ such pairings, all producing the same result, thus 
\begin{align}
	\gamma \: =& \; -\beta \mu\lambda^2
	\frac{6 ! \times \pi }{(48)^2} \int \frac{\dd k_1 \dots \dd k_6}{\omega_1 \dots \omega_6} \delta(k_1 + \dots + k_6)
	\nonumber\\
	&\times\sum_{\sigma_i} \delta(\sigma_1 \omega_1 + \dots + \sigma_{6} \omega_{6})
	(\sigma_1 + \dots + \sigma_6)^2 \frac{1}{\beta^6 (\omega_1 - \mu) \dots (\omega_6 - \mu)}  
	\nonumber\\
	\; =& \lambda^2 \beta^{-5} 
	\frac{120 \times 6! \times \pi}{(48)^2}
	\int \frac{\dd k_1 \dots \dd k_6}{\omega_1 \dots \omega_6} 
	\frac{-\mu}{(\omega_1 - \mu) \dots (\omega_6 - \mu)} 
	\nonumber\\
	&\times\delta(k_1 + \dots + k_6)
	\delta (\omega_1 + \dots + \omega_4 - \omega_5 - \omega_6) \, . 
	\label{eq: rate p=0 formula 2}
\end{align}
This yields eq.~\eqref{eq: r=6 delta > 0.3} after numerical evaluation of the remaining integral.

\bigskip
\underline{$r=4$ and $\delta > 0.3$:} In this case $p=2$.
Our aim is to show that 
\begin{equation}\label{eq: r=4 delta > 0.3}
	\gamma = \gamma_0 \lambda^4 \beta^{-5} \quad \text{for} \quad \mu = -1
\end{equation}
where $\gamma_0$ is a number that depends only on the value of $\delta$ and that can be evaluated explicitly. 

We first need to evaluate $\mathcal J$ up to corrections of order $\mathcal O (\lambda)$. 
From eq.~\eqref{eq: evaluation of the current new variables}, we get 
$$
	\{ \tilde H , N_0 \} = \lambda^2 \{ T_1 H_0 + S_1 V , N_0 \} + \mathcal O (\lambda^3).
$$
Hence we may set $\mathcal J = \{ T_1 H_0 + S_1 V , N_0 \}$, 
with $S_1 = \mathrm{ad}_{G_1}$ and $T_1 = \frac12 \mathrm{ad}_{G_1}\mathrm{ad}_{G_1}$, cfr.~(\ref{eq: first expansion}), 
and where $G_1$ solves $\{ H_0 , G_1\} = V_\perp$, cfr.~(\ref{eq: recursive equations Gn}). 
Hence,
\begin{equation}\label{eq: expression for J for p = 2}
	\mathcal J = \{ \tilde V , N_0 \} 
	\quad \text{with} \quad
	\tilde V = -\frac12 \{G_1 , V + V_\parallel \}
	\quad \text{and} \quad
	\{ H_0 , G_1\} = V_\perp .
\end{equation}
We compute
$$
	G_1 
	= 
	\frac{-i}{16}\int \frac{\dd k_1 \dots \dd k_4}{(\omega_1 \dots \omega_4)^{1/2}} \delta (k_1 + \dots + k_4)
	\sum_{\{ \sigma_i \} \in \perp}  \frac{1}{\sigma_1 \omega_1 + \dots + \sigma_4 \omega_4} a_1^{\sigma_1} \dots a_4^{\sigma_4 } 
$$
where $\{ \sigma_i \} \in \perp$ means the $(\sigma_i)_i$ such that $\sum_i \sigma_i \ne 0$. 
Therefore,
\begin{multline*}
	\tilde V
	 = 
	\frac{i}{2(16)^2} \int \frac{\dd k_1 \dots \dd k_8}{(\omega_1 \dots \omega_8)^{1/2}}\delta (k_1 + \dots + k_4) \delta (k_5 + \dots + k_8) \\
	\times\sum_{\{ \sigma_i \} \in \perp} \left( \sum_{\{ \sigma_j \}} + \sum_{\{ \sigma_j \} \in \parallel} \right)
	\frac{1}{\sigma_1 \omega_1 + \dots + \sigma_4 \omega_4}
	\{ a_1^{\sigma_1} \dots a_4^{\sigma_4} , a_5^{\sigma_5} \dots a_8^{\sigma_8}\}
\end{multline*}
where it is understood that $i = 1, \dots , 4$ for $\{ \sigma_i \}$ and that $j = 5, \dots , 8$ for $\{ \sigma_j \}$. 
To simplify the exposition, we introduce the notation $\{ \sigma_j \} \in \Omega^\parallel$, 
meaning that we sum over all $\{\sigma_j\}$ and that it is counted twice if $\{\sigma_j \} \in \parallel$.
Performing the Poisson bracket yields
\begin{multline*}
	\tilde V
	= 
	\frac{1}{2(16)^2}
	\int \frac{\dd k_1 \dots \dd k_8}{(\omega_1 \dots \omega_8)^{1/2}}\delta (k_1 + \dots + k_4) \delta (k_5 + \dots + k_8) 
	\sum_{i= 1}^4 \sum_{j = 5}^8 \delta(k_i + k_j)\\
	\times \sum_{\{ \sigma_i \} \in \perp} \sum_{\{ \sigma_j \} \in \Omega^\parallel} 
	\frac{ \sigma_i \delta (\sigma_i + \sigma_j) }{\sigma_1 \omega_1 + \dots + \sigma_4 \omega_4}
	a_1^{\sigma_1} \dots \hat a_i^{\sigma_i}  \dots  \hat a_j^{\sigma_j} \dots a_8^{\sigma_8}
\end{multline*}
where $\hat a$ means that this factor is omitted. 
Hence,
\begin{multline*}
	\mathcal J = \{ \tilde V, N_0 \} 
	= 
	\frac{-i}{2(16)^2}
	\int \frac{\dd k_1 \dots \dd k_8}{(\omega_1 \dots \omega_8)^{1/2}}\delta (k_1 + \dots + k_4) \delta (k_5 + \dots + k_8) 
	\sum_{i= 1}^4 \sum_{j = 5}^8 \delta(k_i + k_j)\\
	\times \sum_{\{ \sigma_i \} \in \perp} \sum_{\{ \sigma_j \} \in \Omega^\parallel} 
	\sigma_i \delta (\sigma_i + \sigma_j)
	\frac{ \sigma_1 + \dots + \hat \sigma_i + \dots + \hat \sigma_j + \dots + \sigma_8}{\sigma_1 \omega_1 + \dots + \sigma_4 \omega_4}
	a_1^{\sigma_1} \dots \hat a_i^{\sigma_i}  \dots  \hat a_j^{\sigma_j} \dots a_8^{\sigma_8} .
\end{multline*}

Next, to compute $(\mathrm{ad}_{H_0} + \varepsilon)^{-1} \{ V, N_0 \}$, we use again the expression \eqref{eq: inversion 1}, 
and again only the real part in the fraction featuring in eq.~\eqref{eq: inversion 1} will bring a non-zero contribution. 
We will thus make use of eq.~\eqref{eq: inversion 2} and, for notational simplicity, we will omit the term issuing from the imaginary part: 
\begin{multline*}
	\lim_{\varepsilon\to 0}(\mathrm{ad}_{H_0} + \varepsilon)^{-1}\{ \tilde V , N_0 \} 
	\; = \; 
	\frac{-i \pi}{2(16)^2}
	\int \frac{\dd k_1 \dots \dd k_8}{(\omega_1 \dots \omega_8)^{1/2}}\delta (k_1 + \dots + k_4) \delta (k_5 + \dots + k_8) 
	\sum_{i= 1}^4 \sum_{j = 5}^8 \delta(k_i + k_j)\\
	\times  \sum_{\{ \sigma_i \} \in \perp} \sum_{\{ \sigma_j \} \in \Omega^\parallel} 
	\sigma_i \delta (\sigma_i + \sigma_j)
	\delta (\sigma_1 \omega_1 + \dots + \hat\sigma_i \hat\omega_i + \dots + \hat\sigma_j \hat\omega_j + \dots + \sigma_8 \omega_8)
	\frac{ \sigma_1 + \dots + \hat \sigma_i + \dots + \hat \sigma_j + \dots + \sigma_8}{\sigma_1 \omega_1 + \dots + \sigma_4 \omega_4}\\
	\times  a_1^{\sigma_1} \dots \hat a_i^{\sigma_i}  \dots  \hat a_j^{\sigma_j} \dots a_8^{\sigma_8}
\end{multline*}
Let us simplify this expression. 
By symmetry (changing the labels of the variables), the 16 terms of the sum over $i,j$ all yield the same result, 
hence we may write 
\begin{multline*}
	(\mathrm{ad}_{H_0} + \varepsilon)^{-1}\{ \tilde V , N_0 \} 
	\; = \; 
	\frac{-i 16 \pi}{2(16)^2}
	\int \frac{\dd k_1 \dots \dd k_8}{(\omega_1 \dots \omega_8)^{1/2}}\delta (k_1 + \dots + k_4) \delta (k_5 + \dots + k_8) \delta(k_4 + k_8) \\
	\times  \sum_{\{ \sigma_i \} \in \perp} \sum_{\{ \sigma_j \} \in \Omega^\parallel} 
	\sigma_4 \delta (\sigma_4 + \sigma_8)
	\delta (\sigma_1 \omega_1 + \dots + \sigma_3 \omega_3 + \sigma_5 \omega_5 + \dots + \sigma_7 \omega_7)
	\frac{ \sigma_1 + \dots + \sigma_3 + \sigma_5 + \dots + \sigma_7}{\sigma_1 \omega_1 + \dots + \sigma_4 \omega_4}\\
	\times  a_1^{\sigma_1} \dots a_3^{\sigma_3} a_5^{\sigma_5}\dots a_7^{\sigma_7}
\end{multline*}
Next, thanks to the energy constraint and the sum $\sigma_1 + \dots + \sigma_3 + \sigma_5 + \dots + \sigma_7$, 
only monomials with $++++--$ or $++----$ do yield a non-zero contribution. 
This will allow an explicit summation over the $\sigma$ configurations.
Let us compute the $++++--$ term (the term $++----$ will be obtained by reversing the signs of all $\sigma$).
We identify all configurations that yield a non-zero contribution as: 

$$
	\begin{array}{c|cc}
	\{ \sigma_i \} \in \perp &  \{ \sigma_i \} & \\
	\hline
	++++  & +--- \text{ or } -+-- \text{ or } --+ - & \in \perp\\
	+++-  & +--+ \text{ or } -+-+ \text{ or } --++ &\in \parallel\\
	++-+ \text{ or } +-++ \text{ or } -+++ & ++-- \text{ or } +-+- \text{ or } -++- & \in \parallel\\
	+--- \text{ or } -+-- \text{ or } --+- & ++++ & \in \perp
	\end{array}
$$
where we have taken into account that the last $\pm$ in each configuration in the first column must be paired with the last $\mp$ in the second column.  
Elements in $\parallel$ must be counted twice (in the second column). 
In all cases, we get  $\sigma_1 + \dots + \sigma_3 + \sigma_5 + \dots + \sigma_7 = 2$.
We arrive at
\begin{align*}
	&
	(\mathrm{ad}_{H_0} + \varepsilon)^{-1}\{ \tilde V , N_0 \} 
	\; = \; 
	\frac{-i \pi}{32}
	\int \frac{\dd k_1 \dots \dd k_8}{(\omega_1 \dots \omega_8)^{1/2}}\delta (k_1 + \dots + k_4) \delta (k_5 + \dots + k_8) \delta(k_4 + k_8) \\
	& \qquad \times 
	\bigg\{ \bigg[ 
	3 \times \delta(\omega_1 + \omega_2 + \omega_3 + \omega_5 - \omega_6 - \omega_7) 
	\frac{2}{\omega_1 + \omega_2 + \omega_3 + \omega_4} a_1^+ a_2^+ a_3^+ a_5^+ a_6^-a_7^- \\
	&\qquad
	-2\times 3 \times \delta(\omega_1 + \omega_2 + \omega_3 + \omega_5 - \omega_6 - \omega_7) 
	\frac{2}{\omega_1 + \omega_2 + \omega_3 - \omega_4} a_1^+ a_2^+ a_3^+ a_5^+ a_6^-a_7^- \\
	&\qquad
	+ 2 \times 9 \times \delta(\omega_1 + \omega_2 - \omega_3 + \omega_5 + \omega_6 - \omega_7) 
	\frac{2}{\omega_1 + \omega_2 - \omega_3 + \omega_4} a_1^+ a_2^+ a_3^- a_5^+ a_6^+ a_7^- \\
	&\qquad
	- 3 \times \delta(\omega_1 - \omega_2 - \omega_3 + \omega_5 + \omega_6 + \omega_7) 
	\frac{2}{\omega_1 - \omega_2 - \omega_3 - \omega_4} a_1^+ a_2^- a_3^- a_5^+ a_6^+ a_7^+
	\bigg]\\
	&\qquad -\\
	&\qquad 
	\bigg[
	3 \times \delta(\omega_1 + \omega_2 + \omega_3 + \omega_5 - \omega_6 - \omega_7) 
	\frac{2}{\omega_1 + \omega_2 + \omega_3 + \omega_4} a_1^- a_2^- a_3^- a_5^- a_6^+ a_7^+ \\
	&\qquad
	-2\times 3 \times \delta(\omega_1 + \omega_2 + \omega_3 + \omega_5 - \omega_6 - \omega_7) 
	\frac{2}{\omega_1 + \omega_2 + \omega_3 - \omega_4} a_1^- a_2^- a_3^- a_5^- a_6^+ a_7^+ \\
	&\qquad
	+ 2 \times 9 \times \delta(\omega_1 + \omega_2 - \omega_3 + \omega_5 + \omega_6 - \omega_7) 
	\frac{2}{\omega_1 + \omega_2 - \omega_3 + \omega_4} a_1^- a_2^- a_3^+ a_5^- a_6^- a_7^+ \\
	&\qquad
	- 3 \times \delta(\omega_1 - \omega_2 - \omega_3 + \omega_5 + \omega_6 + \omega_7) 
	\frac{2}{\omega_1 - \omega_2 - \omega_3 - \omega_4} a_1^- a_2^+ a_3^+ a_5^- a_6^- a_7^-
	\bigg] \bigg\}
\end{align*}
Let us then perform the integration over $k_4,k_8$:
\begin{align*}
	&\int \frac{\dd k_4 \dd k_8}{(\omega_4 \omega_8)^{1/2}} 
	\frac{\delta(k_1 + \dots + k_4) \delta (k_5 + \dots + k_8)\delta(k_4 +k_8)}{\sigma_1 \omega_1 + \sigma_2 \omega_2 + \sigma_3 \omega_3 + \sigma_4 \omega_4} \\
	&\; = \; 
	\frac{\delta (k_1 + k_2 + k_3 + k_5 + k_6 + k_7)}{\omega_0 (\sigma_1\omega_1 + \sigma_2\omega_2 + \sigma_3\omega_3 + \sigma_4 \omega_0)}
	\qquad \text{with}\qquad \omega_0 = \omega (k_1 + k_2 + k_3).
\end{align*}
By changing the labels of the variables, and bringing in front an overall factor $6$, we get 
\begin{align}
	&
	(\mathrm{ad}_{H_0} + \varepsilon)^{-1}\{ \tilde V , N_0 \} 
	\; = \; 
	\frac{-i 6 \pi}{32}
	\int \frac{\dd k_1 \dots \dd k_6}{(\omega_1 \dots \omega_6)^{1/2}}
	\delta(k_1 + \dots + k_6) \delta (\omega_1 + \omega_2 + \omega_3 + \omega_4 - \omega_5 - \omega_6)\nonumber\\
	& \times \bigg[
	\frac{1}{\omega_0 (\omega_1 + \omega_2 + \omega_3 + \omega_0)}
	- \frac{2}{\omega_0 (\omega_1 + \omega_2 + \omega_3 - \omega_0)}
	+ \frac{6}{\omega_0 (\omega_1 + \omega_2 - \omega_5 + \omega_0)}
	- \frac{1}{\omega_0 (\omega_1 - \omega_5 - \omega_6 - \omega_0)}
	\bigg]\nonumber\\
	&\times  \big( a_1^+ a_2^+ a_3^+ a_4^+ a_5^- a_6^- - a_1^- a_2^- a_3^- a_4^- a_5^+ a_6^+ \big) . 
	\label{eq: expression with bracket}
\end{align}
Here we used the convention that, in the expression $\omega_0 (\omega_a \pm \omega_b \pm \omega_c \pm \omega_0)$, we have $\omega_0 = \omega(k_a + k_b + k_c)$.

Finally, we compute $\langle \{ \tilde V , N_0 \} (\mathrm{ad}_{H_0} + \varepsilon)^{-1}\{ \tilde V , N_0 \} \rangle$. 
Let us give a name to the expression $[\dots]$ in eq.~\eqref{eq: expression with bracket}:
$$
	\mathcal F(k_1, \dots , k_6) = [\dots].
$$
To perform the Gaussian pairings, we partially symmetrize $\mathcal F$ (additively), so that it is symmetric under the exchanges of the variables $1,\dots,4$ and $5,6$.
We denote by $S\mathcal F$ the partial symmetrization of $\mathcal F$. 
Because of the energy constraint, we realize again that pairings must be between variables with indices $1, \dots , 6$ on the one hand, 
and $7, \dots , 12$ on the other hand. 
We get
\begin{align*}
	& \langle \{ \tilde V_0 , N_0 \} (\mathrm{ad}_{H_0} + \varepsilon)^{-1}\{ \tilde V , N_0 \} \rangle
	\; = \; 
	2 \times (3/16)^2 \times  \pi \times 2 \times 4! \times \delta (0)\\
	& \qquad \times \int \frac{\dd k_1 \dots \dd k_6}{(\omega_1 \dots \omega_6)}
	\delta(k_1 + \dots + k_6) \delta (\omega_1 + \omega_2 + \omega_3 + \omega_4 - \omega_5 - \omega_6)
	\big( S\mathcal F (k_1, \dots , k_6)\big)^2
	\frac{1}{\beta^6 (\omega_1 - \mu) \dots (\omega_6 - \mu)}
\end{align*}
Hence,
\begin{align}
	& \gamma
	\; = \; 
	\lambda^4\beta^{-5} \times
	4 \times (3/16)^2 \times 4! \times \pi  \times (-\mu) \nonumber\\
	&\qquad \times \int \frac{\dd k_1 \dots \dd k_6}{(\omega_1 \dots \omega_6) (\omega_1 - \mu) \dots (\omega_6 - \mu)}
	\delta(k_1 + \dots + k_6) \delta (\omega_1 + \omega_2 + \omega_3 + \omega_4 - \omega_5 - \omega_6) \big( S\mathcal F (k_1, \dots , k_6)\big)^2.
\end{align}
This yields \eqref{eq: r=4 delta > 0.3}.

\bigskip
\underline{$r=6$ and $0.255 < \delta < 0.3$:} In this case $p=2$. 
Our aim is to show that 
\begin{equation}\label{eq:r6deltalt0.3}
	\gamma = \gamma_0 \lambda^4 \beta^{-9} \quad \text{for} \quad \mu = -1
\end{equation}
where $\gamma_0$ is a number that depends only on the value of $\delta$ and that can be evaluated explicitly.

To a large extend, the computation parallels the computation for the case $r=4,\delta>0.3$, and we omit intermediate steps when possible. 
The expression for $\mathcal J$ is still given by \eqref{eq: expression for J for p = 2}
Next, the pre-factor $1/r 2^{r/2}$ needs to be changed from $16$ to $48$, 
the indices $1,\dots ,4$ and $5,\dots, 8$ become respectively $1,\dots,6$ and $7,\dots,12$, and 
finally the summation over $i,j$ brings an overall factor $6^2$ instead $4^2$. 
Hence we arrive at 
\begin{multline*}
	(\mathrm{ad}_{H_0} + \varepsilon)^{-1}\{ \tilde V , N_0 \} 
	\; = \; 
	\frac{-i 6^2 \pi}{2(48)^2}
	\int \frac{\dd k_1 \dots \dd k_{12}}{(\omega_1 \dots \omega_{12})^{1/2}}\delta (k_1 + \dots + k_6) \delta (k_7 + \dots + k_{12}) \delta(k_6 + k_{12}) \\
	\times \sum_{\{ \sigma_i \} \in \perp} \sum_{\{ \sigma_j \} \in \Omega^\parallel} 
	\sigma_6 \delta (\sigma_{6} + \sigma_{12})
	\delta (\sigma_1 \omega_1 + \dots + \sigma_5 \omega_5 + \sigma_7 \omega_7 + \dots + \sigma_{11} \omega_{11})
	\frac{ \sigma_1 + \dots + \sigma_5 + \sigma_7 + \dots + \sigma_{11}}{\sigma_1 \omega_1 + \dots + \sigma_6 \omega_6}\\
	\times a_1^{\sigma_1} \dots a_5^{\sigma_5} a_7^{\sigma_7}\dots a_{11}^{\sigma_{11}}.
\end{multline*}
Again, thanks to the summation $\sigma_1 + \dots + \sigma_5 + \sigma_7 + \dots + \sigma_{11}$, 
and thanks to the energy constraint, only the terms $++++++----$ and $++++------$ do yield a non-zero contribution. 
Let us again compute the term $++++++----$. 
For this, we list all possibilities, remembering that we pair the last $\pm$ in the 1st column below with the last $\mp$ in the second column. 
This time however, we will not explicitly write configurations that differ only by permutations.
Instead we will indicate the number of such terms (with the sign), remembering that terms in $\parallel$ are counted twice: 
$$
\begin{array}{c|cc|c}
	\{ \sigma_i \} \in \perp &  \{ \sigma_i \} && \text{counting} \\
	\hline
	+++++ +  & + ---- -   & \in \perp     & 5\\
	+++++ -  & + ---- +   & \in \perp     & -5\\
	-++++ +  & ++ --- -   & \in \perp     & 5 \times 10\\
	-++++ -  & ++ --- +   & \in \parallel & -2 \times 5 \times 10\\
	--+++ +  & +++ -- -   & \in \parallel & 2 \times 10 \times 10\\
	---++ -  & ++++ - +   & \in \perp     & - 10 \times 5\\
	----+ +  & +++++  -   & \in \perp     & 5\\
	----+ -  & +++++  +   & \in \perp     & -5
\end{array}
$$
We notice that there is an overall factor $5$ for all these terms, and that the summation $\sigma_1 + \dots + \sigma_5 + \sigma_7 + \dots + \sigma_{11} = 2$ always. 
Hence we get
\begin{align*}
	&(\mathrm{ad}_{H_0} + \varepsilon)^{-1}\{ \tilde V , N_0 \} 
	\; = \; 
	\frac{-i 2 \times 5 \times 6^2 \times \pi}{2(48)^2}
	\int \frac{\dd k_1 \dots \dd k_{12}}{(\omega_1 \dots \omega_{12})^{1/2}}\delta (k_1 + \dots + k_6) \delta (k_7 + \dots + k_{12}) \delta(k_6 + k_{12}) \\
	&\times \bigg(\bigg[\frac{\delta (\omega_1 + \omega_2 + \omega_3 + \omega_4 + \omega_5 + \omega_7 - \omega_8 - \omega_9 - \omega_{10} - \omega_{11})}
	{\omega_1 + \omega_2 + \omega_3 + \omega_4 + \omega_5 + \omega_6} 
	a_1^+ a_2^+ a_3^+ a_4^+ a_5^+ a_7^+ a_8^- a_9^- a_{10}^- a_{11}^-\\
	&-\frac{\delta (\omega_1 + \omega_2 + \omega_3 + \omega_4 + \omega_5 + \omega_7 - \omega_8 - \omega_9 - \omega_{10} - \omega_{11})}
	{\omega_1 + \omega_2 + \omega_3 + \omega_4 + \omega_5 - \omega_6} 
	a_1^+ a_2^+ a_3^+ a_4^+ a_5^+ a_7^+ a_8^- a_9^- a_{10}^- a_{11}^-\\
	&+10\frac{\delta (\omega_1 + \omega_2 + \omega_3 + \omega_4 - \omega_5 + \omega_7 + \omega_8 - \omega_9 - \omega_{10} - \omega_{11})}
	{\omega_1 + \omega_2 + \omega_3 + \omega_4 - \omega_5 + \omega_6} 
	a_1^+ a_2^+ a_3^+ a_4^+ a_5^- a_7^+ a_8^+ a_9^- a_{10}^- a_{11}^-\\
	&-20\frac{\delta (\omega_1 + \omega_2 + \omega_3 + \omega_4 - \omega_5 + \omega_7 + \omega_8 - \omega_9 - \omega_{10} - \omega_{11})}
	{\omega_1 + \omega_2 + \omega_3 + \omega_4 - \omega_5 - \omega_6} 
	a_1^+ a_2^+ a_3^+ a_4^+ a_5^- a_7^+ a_8^+ a_9^- a_{10}^- a_{11}^-\\
	&+40\frac{\delta (\omega_1 + \omega_2 + \omega_3 - \omega_4 - \omega_5 + \omega_7 + \omega_8 + \omega_9 - \omega_{10} - \omega_{11})}
	{\omega_1 + \omega_2 + \omega_3 - \omega_4 - \omega_5 + \omega_6} 
	a_1^+ a_2^+ a_3^+ a_4^- a_5^- a_7^+ a_8^+ a_9^+ a_{10}^- a_{11}^-\\
	&-10\frac{\delta (\omega_1 + \omega_2 - \omega_3 - \omega_4 - \omega_5 + \omega_7 + \omega_8 + \omega_9 + \omega_{10} - \omega_{11})}
	{\omega_1 + \omega_2 - \omega_3 - \omega_4 - \omega_5 - \omega_6} 
	a_1^+ a_2^+ a_3^- a_4^- a_5^- a_7^+ a_8^+ a_9^+ a_{10}^+ a_{11}^-\\
	&+\frac{\delta (\omega_1 - \omega_2 - \omega_3 - \omega_4 - \omega_5 + \omega_7 + \omega_8 + \omega_9 + \omega_{10} + \omega_{11})}
	{\omega_1 - \omega_2 - \omega_3 - \omega_4 - \omega_5 + \omega_6} 
	a_1^+ a_2^- a_3^- a_4^- a_5^- a_7^+ a_8^+ a_9^+ a_{10}^+ a_{11}^+\\
	&-\frac{\delta (\omega_1 - \omega_2 - \omega_3 - \omega_4 - \omega_5 + \omega_7 + \omega_8 + \omega_9 + \omega_{10} + \omega_{11})}
	{\omega_1 - \omega_2 - \omega_3 - \omega_4 - \omega_5 - \omega_6} 
	a_1^+ a_2^- a_3^- a_4^- a_5^- a_7^+ a_8^+ a_9^+ a_{10}^+ a_{11}^+ \bigg]\\
	&- [\dots] \bigg)
\end{align*}
where the expression in $[\dots]$ is the same as the previous with all $a^\pm$ changed to $a^\mp$.
Next, we integrate over $k_6,k_{12}$: 
\begin{align*}
	& \int \frac{\dd k_6 \dd k_{12}}{(\omega_6 \omega_{12})^{1/2}}
	\frac{\delta (k_1 + \dots + k_6) \delta (k_7 + \dots + k_{12}) \delta(k_6 + k_{12})}{\sigma_1 \omega_1 + \dots + \sigma_6 \omega_6} \\
	&\; = \;  
	\frac{\delta (k_1 + \dots + k_5 + k_7 + \dots + k_{11})}{\omega_0 (\sigma_1 \omega_1 + \dots + \sigma_5 \omega_5 + \sigma_0 \omega_0)}
	\quad \text{with}\quad 
	\omega_0 = \omega(k_1 + \dots + k_5) \, .
\end{align*}
Hence, by changing the labels of the variables, we obtain 
\begin{align*}
	&(\mathrm{ad}_{H_0} + \varepsilon)^{-1}\{ \tilde V_0 , N_0 \} 
	\; = \; 
	- i \frac{5 \pi}{64} \int \frac{\dd k_1 \dots \dd k_{10}}{(\omega_1 \dots \omega_{10})^{1/2}} 
	\delta(k_1 + \dots + k_{10})
	\delta(\omega_1 + \dots + \omega_6 - \omega_7 - \dots - \omega_{10}) \\
	&\times \bigg[ 
	       \frac{1}{\omega_0 (\omega_1 + \omega_2 + \omega_3 + \omega_4 + \omega_5 + \omega_0)}
	  -    \frac{1}{\omega_0 (\omega_1 + \omega_2 + \omega_3 + \omega_4 + \omega_5 - \omega_0)} \\
	& +   \frac{10}{\omega_0 (\omega_1 + \omega_2 + \omega_3 + \omega_4 - \omega_7 + \omega_0)} 
	  -   \frac{20}{\omega_0 (\omega_1 + \omega_2 + \omega_3 + \omega_4 - \omega_7 - \omega_0)} \\
	& +   \frac{40}{\omega_0 (\omega_1 + \omega_2 + \omega_3 - \omega_7 - \omega_8 + \omega_0)} 
	  -   \frac{10}{\omega_0 (\omega_1 + \omega_2 - \omega_7 - \omega_8 - \omega_9 - \omega_0)} \\
	& +    \frac{1}{\omega_0 (\omega_1 - \omega_7 - \omega_8 - \omega_9 - \omega_{10} + \omega_0)} 
	  -    \frac{1}{\omega_0 (\omega_1 - \omega_7 - \omega_8 - \omega_9 - \omega_{10} - \omega_0)} 
	\bigg]\nonumber\\
	&\times \big( a_1^+ a_2^+ a_3^+ a_4^+ a_5^+ a_6^+ a_7^- a_8^- a_9^- a_{10}^- -  a_1^- a_2^- a_3^- a_4^- a_5^- a_6^- a_7^+ a_8^+ a_9^+ a_{10}^+\big).
\end{align*}
Again, we denote the expression in $[\dots]$ by $\mathcal F$ and we consider the partial symmetrization $S \mathcal F$
with respect to the variables $1, \dots , 6$ and $7, \dots , 10$. 

Finally, we evaluate $\langle \{ \tilde V_0 , N_0 \}  (\mathrm{ad}_{H_0} + \varepsilon)^{-1}\{ \tilde V_0 , N_0 \}  \rangle$ and perform Gaussian pairings. 
At the variance of the case $r=4$ and $\delta > 0.3$ treated above, there are now two distinct possibilities. 
First, as before, we may pair each of the variables $1, \dots , 10$ with a variable $11, \dots , 20$. 
Second, and this is new, we may pair two variables of the group $1, \dots, 10$ among them, and two variables of the group $11, \dots , 20$ among them, 
and then pair the variables of the first group with variables of the second group. 
It is not possible to pair 4 or more variables from a same group, otherwise the energy constraint cannot be realized. 
So we decompose
$$
	\langle \{ \tilde V_0 , N_0 \}  (\mathrm{ad}_{H_0} + \varepsilon)^{-1}\{ \tilde V_0 , N_0 \}  \rangle \; = \; I_1 + I_2
$$
and compute separately each term. 

For $I_1$, the computation is as previously. 
There are $2 \times 6! \times 4!$ pairings, hence we get 
\begin{align*}
	I_1 \; =& \; 
	2 \times 6 ! \times 4 ! \times (5/64)^2 \times \pi \times \delta (0)
	\int \frac{\dd k_1 \dots \dd k_{10}}{\omega_1 \dots \omega_{10}}
	\frac{1}{\beta^{10} (\omega_1 - \mu) \dots (\omega_{10} - \mu)} \\
	&\times \delta (k_1 + \dots + k_{10})
	\delta (\omega_1 + \dots + \omega_6 - \omega_7 - \dots - \omega_{10})
	(S\mathcal F)^2
\end{align*}
For $I_2$, we first do the two internal pairings. 
This amounts to replace the function $\{ \tilde V , N_0 \}$ by a function $g$ where the internal pairing is done. 
There are $6\times 4 = 24$ ways of making this pairing and by symmetry, we get
\begin{align*}
	(\mathcal L_0 + \varepsilon)^{-1} g 
	\; =& \; 
	- i \frac{5 \pi }{64} \times 24 
	\int \frac{\dd k_1 \dots \dd k_8}{(\omega_1 \dots \omega_8)^{1/2}} 
	\delta (k_1 + \dots + k_8) \delta (\omega_1 + \dots + \omega_5 - \omega_6 - \omega_7 - \omega_8)\\
	&\times \int \frac{\dd k}{\omega (k) \beta (\omega (k) - \mu)} S\mathcal F (k , k_1, \dots, , k_5, -k, k_6,k_7,k_8)\\
	&\times (a_1^+a_2^+a_3^+a_4^+a_5^+ a_6^- a_7^- a^-_8 - a_1^- a_2^- a_3^- a_4^- a_5^- a_6^+ a_7^- a_8^- ) 
\end{align*}
and we introduce the notation
$$
	G(k_1, \dots , k_8) \; = \; \int \frac{\dd k}{\omega (k) \beta (\omega (k) - \mu)} S\mathcal F (k , k_1, \dots, , k_5, -k, k_6,k_7,k_8) \, 
$$
which is still symmetric in the variables $1, \dots , 5$ and $6,7,8$. 
Hence, we obtain 
\begin{align*}
	I_2 \; = &\; 
	2 \times (15/8)^2 \times 5! \times 3! \times \pi \times \delta (0)
	\int \frac{\dd k_1 \dots \dd k_8}{\omega_1 \dots \omega_8} \frac{1}{\beta^8 (\omega_1 - \mu) \dots (\omega_8 - \mu)} \\
	&\times \delta (k_1 + \dots + k_8) \delta (\omega_1 + \dots + \omega_5 - \omega_6 - \omega_7 - \omega_8) G^2 \, .
\end{align*}
Finally, 
$$
	\gamma = - \beta \mu \lambda^4 \frac{(I_1 + I_2)}{\delta (0)}. 
$$
This yields eq.~\eqref{eq:r6deltalt0.3}.

\section{Numerical procedure}

All data points are generated by directly simulating the dynamics for the Hamiltonian $H$ in eq.~(\ref{eq: hamiltonian})
with $L=1024$ and periodic boundary conditions.
The numerical scheme is a standard Str\"omer--Verlet algorithm with a time step $\Delta t = 0.1$.
For large $\lambda$, where one does not need to follow the dynamics on very long time scales, 
we have checked that changing $L$ and $\Delta t$ only produces marginal differences.

Let us fix the parameters $\lambda,\delta$ of the Hamiltonian $H$.
Initially, we fix $\beta > 0$ and $\mu = -1$ that determine the initial state, and we fix the value of each phonon mode to be
\begin{equation}\label{eq: initial state SM}
	a(k) = \sqrt{W(k)} \ed^{i \varphi(k)} ,
	\quad W(k) = \frac{1}{\beta (\omega (k) - \mu)}, 
	\quad 
	\varphi (k) \text{ independent and uniformly distributed in } [0,2 \pi).
\end{equation}
A similar kind of initial state (with different choices for $W$) is used in \cite{mendl_et_al_2016}. 
The data are averaged over $250-4000$ initial configurations, corresponding to different realizations of $\varphi_k$. 

\begin{figure}[h]
    \centering
    \includegraphics[draft=false,height = 4.5cm,width = 5.8cm]{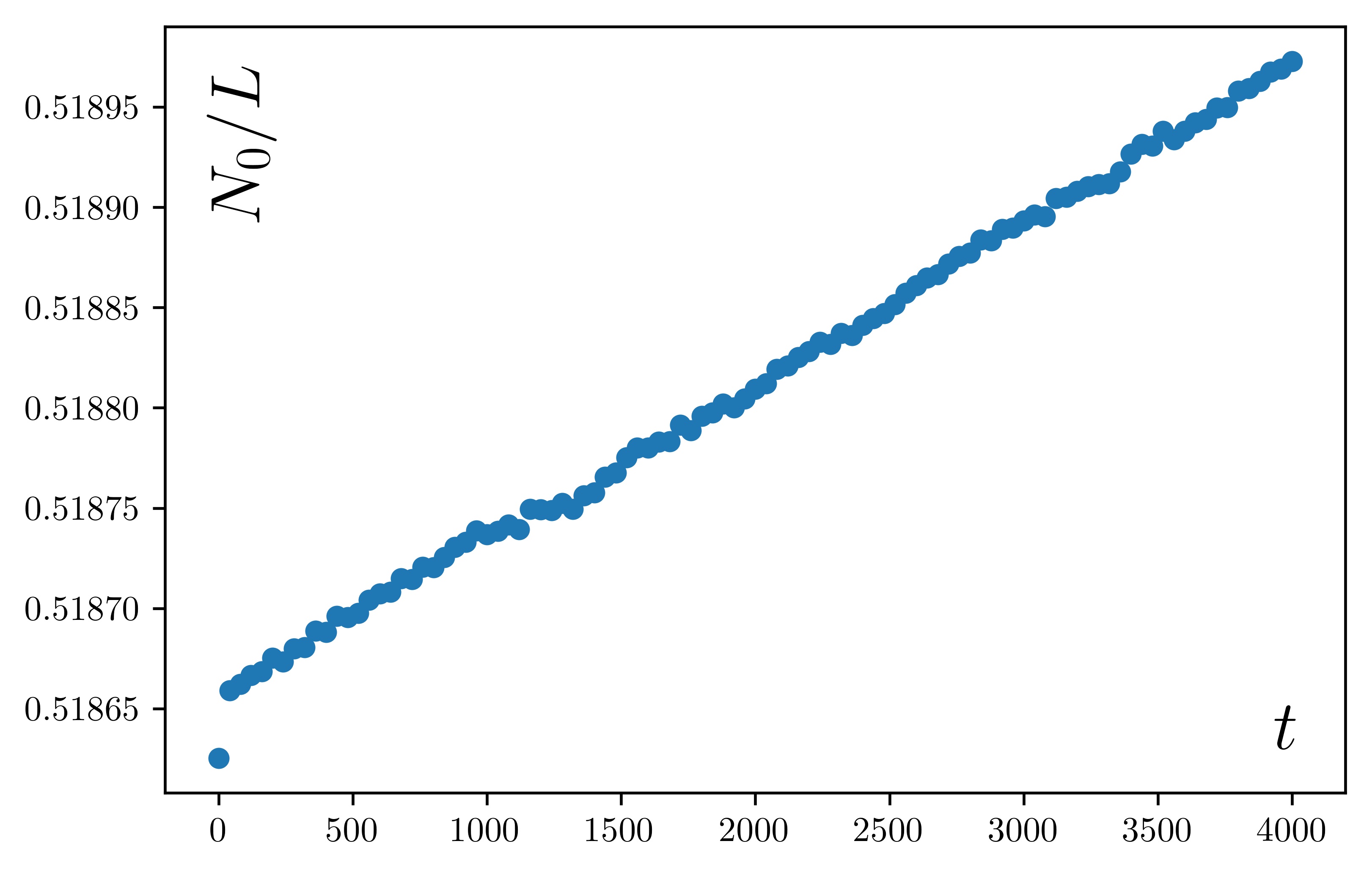}
	\includegraphics[draft=false,height = 4.5cm,width = 5.8cm]{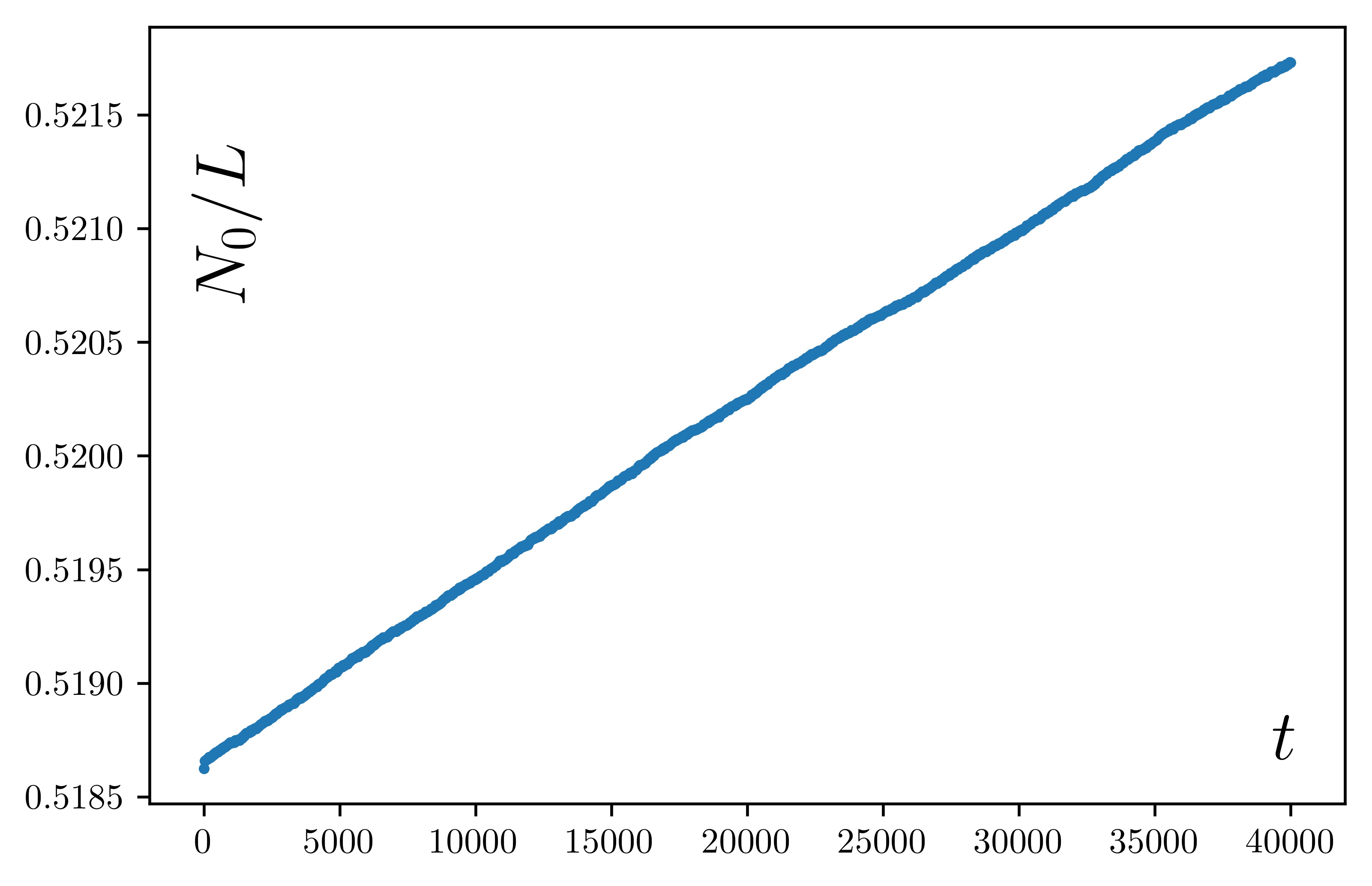}
	\includegraphics[draft=false,height = 4.5cm,width = 5.8cm]{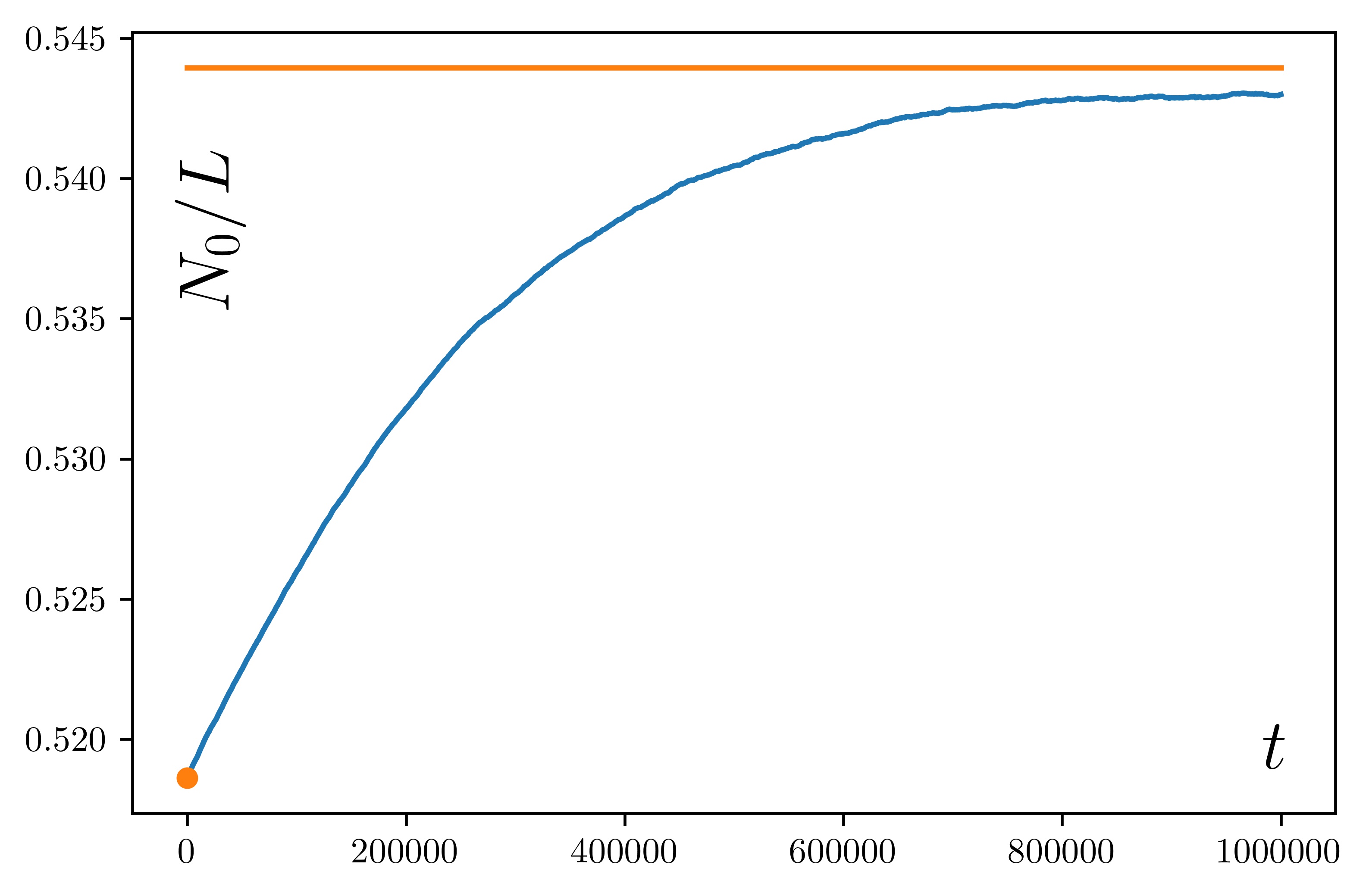}
    \caption{$N_0/L$ as a function of time for $r=6$, $\delta = 0.35$ and $\lambda = 10^{-3}$ on three different time scales.
    Average over more than 2000 initial configurations. 
    Left panel: $0 \le t \le 4 \times 10^{3}$. 
    We observe a leap between $t=0$ and $t=40$, corresponding presumably to the stage where the system moves to the pre-thermal state.
    Middle panel: $0 \le t \le 4 \times 10^4$. 
    The value of $N_0/L$ increases approximately linearly. 
    The slope is taken as the value for $\gamma$ to obtain the corresponding point on Fig.~\ref{fig: r=6 delta=0.35} in the main text.
    Right panel: $0 \le t \le 10^6$. $N_0/L$ reaches eventually its thermal value $(N_0 / L)_{th} \simeq 0.54$ (computed at $\lambda = 0$) (orange). }
    \label{fig: three different focuses}
\end{figure}

\begin{figure}[h]
    \centering
    \includegraphics[draft=false,height = 5.5cm,width = 7.5cm]{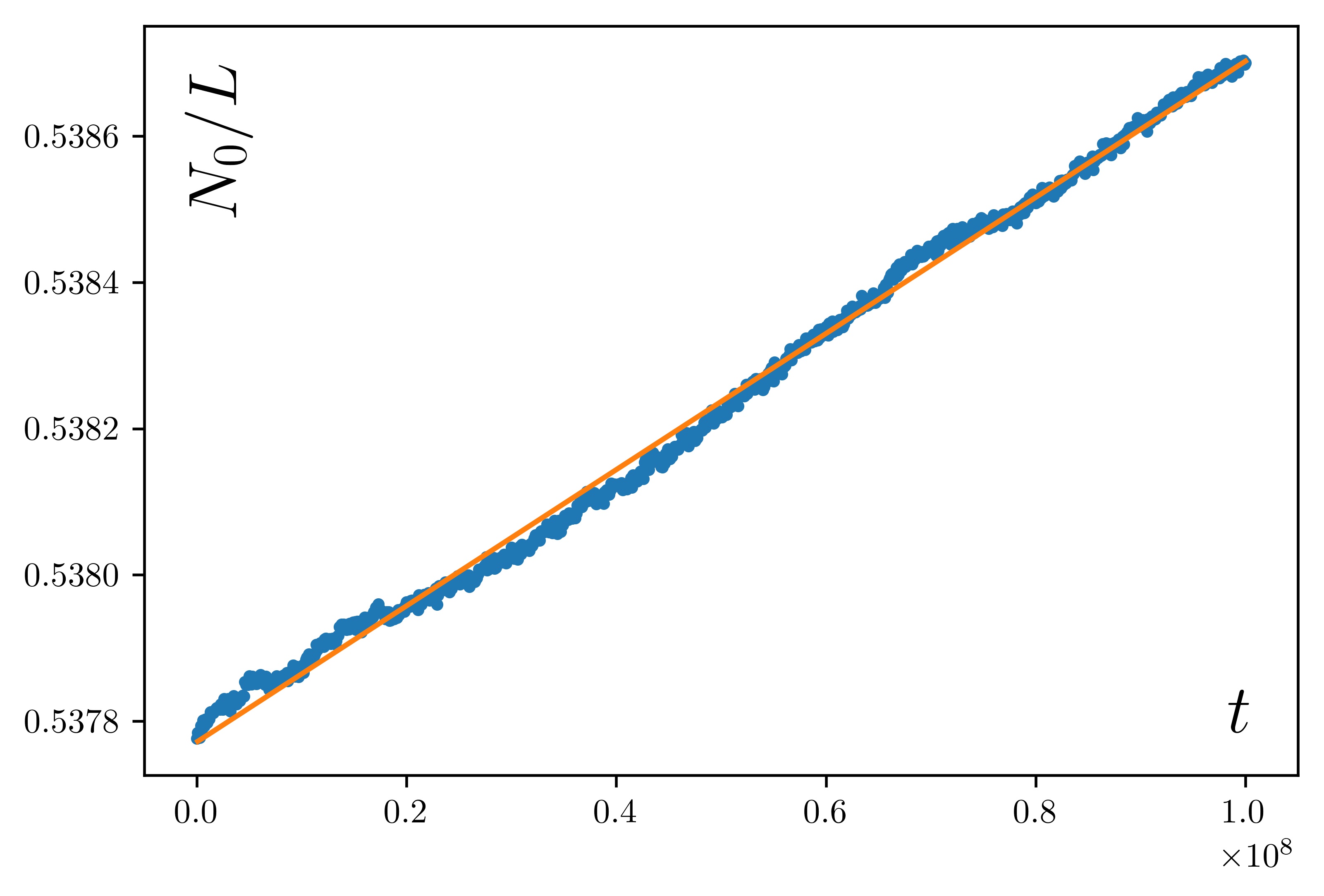}
	\includegraphics[draft=false,height = 5.5cm,width = 7.5cm]{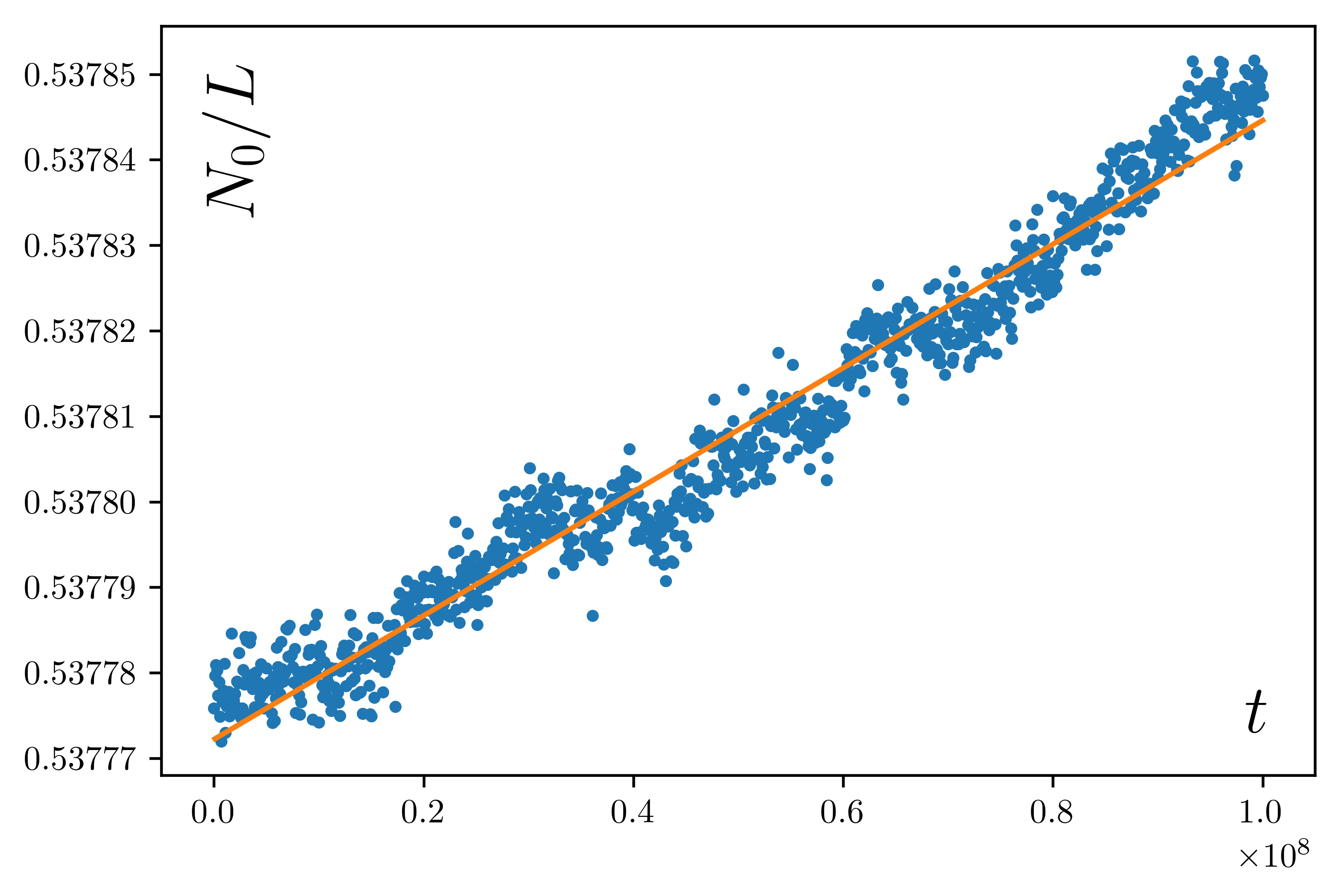}
    \caption{$N_0/L$ as a function of time for $r=4$, $\delta = 0.45$ and $\lambda = 10^{-3}$ (left panel) and $\lambda = 5 \times 10^{-4}$ (right panel). 
    The rate $\gamma$ is determined by a mean square fit (orange). 
    Average over more than 250 initial configurations. }
    \label{fig: smaller lambda}
\end{figure}

Starting from the initial state \eqref{eq: initial state SM}, we expect that the pre-thermal state is reached on very short times, 
and this seems to be indeed the case, see the left panel on Fig.~\ref{fig: three different focuses}.
Next, to measure the rate $\gamma$, we measure how $N_0 / L$ evolves with time, and we observe that the evolution is first approximately linear, 
see the middle panel on Fig.~\ref{fig: three different focuses}. We identify the slope of this linear piece with $\gamma (\beta,\lambda,\delta)$. 
This should become exact in the limit $\lambda \to 0$ that we investigate. 
For large $\lambda$, there is some arbitrariness in determining the time interval where the evolution is approximately linear. 
However, for smaller values of $\lambda$, 
this interval simply corresponds to the longest time on which one is reasonably able to run the simulations and perform sufficient averaging ($t=10^8$).
See Fig.~\ref{fig: smaller lambda}.
Finally, the value of $N_0/L$ reaches its equilibrium value on longer time scales, see the right panel on Fig.~\ref{fig: three different focuses}.

\end{document}